\documentclass[preprint,superscriptaddress,amsmath,amssymb,floatfix, longbibliography, nofootinbib]{revtex4-1}
\usepackage{graphicx}
\usepackage{float}
\usepackage{latexsym,amsmath,amssymb,bm,euscript,multirow,makecell}
\usepackage{color,esvect}
\usepackage{wasysym}
\usepackage{epstopdf}
    \usepackage[colorlinks=true,linkcolor=blue,citecolor=blue,pdfencoding=auto, psdextra,urlcolor=blue]{hyperref}
\usepackage{type1cm}
\usepackage{appendix} 

\usepackage{booktabs,multirow}

\newcommand{\tj}[6]{ \begin{pmatrix}
   #1 & #2 & #3 \\
   #4 & #5 & #6 
  \end{pmatrix}} 

\newcommand{\sj}[6]{\left\{ \begin{matrix}
   #1 & #2 & #3 \\
   #4 & #5 & #6 
  \end{matrix} \right\} }

\renewcommand{\vec}[1]{\bm{#1}}

\newcommand{\bDelta}{\boldsymbol{\Delta}}

\begin{document}

\title{Free real scalar CFT on fuzzy sphere: spectrum, algebra and wavefunction ansatz}
\author{Yin-Chen He}
\affiliation{C. N. Yang Institute for Theoretical Physics, Stony Brook University, Stony Brook, NY 11794-3840}
\affiliation{Perimeter Institute for Theoretical Physics, Waterloo, Ontario N2L 2Y5, Canada}

\begin{abstract}

We introduce a simple model to realize the free real scalar CFT on the fuzzy sphere. The model is structurally similar to the original model that realizes the 3D Ising CFT on the fuzzy sphere. Owing to the shift symmetry of the free scalar, the free scalar CFT fixed point in our model can be accessed with only a single tuning parameter—the conformal coupling. We numerically demonstrate that our model correctly reproduces the operator spectrum, correlation functions, and, crucially, the harmonic oscillator algebra of the real scalar CFT. We also examine the fuzzy sphere algebra, a generalization of the Girvin–MacDonald–Platzman algebra, and discuss its potential implications for defining quantum field theories on non-commutative geometries. Finally, we propose wavefunction ansatz for the ground states of both the free scalar and Ising CFTs, which exhibit remarkable agreement with the CFT  ground state wavefunctions of the fuzzy sphere model. For instance, the wavefunction overlap between our Ising ansatz and the Ising CFT ground state exceeds $0.99$ for $N=28$ fermions, suggesting a promising direction for tackling the 3D Ising CFT.

\end{abstract}

\maketitle

\tableofcontents

\section{Introduction}

Recently, the fuzzy sphere regularization has emerged as a powerful tool for studying 3D conformal field theories (CFTs)~\cite{ZHHHH2022}. The idea is to study 3D CFTs on the cylinder geometry, $S^2 \times \mathbb{R}$, by making use of a fuzzy (non-commutative) sphere~\cite{madore1992fuzzy}. Physically, it is realized by fermions on the lowest spherical Landau levels generated by a magnetic monopole~\cite{Sphere_LL_Haldane}. The fuzzy sphere approach has been successfully applied to a variety of theories, including the 3D Ising CFT~\cite{ZHHHH2022,hu2023operator,Han2023Conformal,hu2024F,Hofmann2024,fardelli2024,fan2024,voinea2024,Lauchli_CPT}, $O(N)$ Wilson-Fisher theories~\cite{Han2023Conformal}, the $SO(5)$ deconfined phase transition~\cite{zhou2023so5,PhysRevLett.132.246503,PhysRevB.110.125153}, symplectic gauge theories~\cite{zhou2024newseries3dcfts}, the 3-state Potts transition~\cite{SYang2025}, Yang-Lee non-unitary CFT~\cite{Fan_YangLee,Xin_YangLee,Miro_LeeYang}, and defect CFTs~\cite{hu2023defect,Zhou2024gfunction,Zhou2025surface,fuzzyHemi,Cuomo2024}. More importantly, thanks to the cylinder geometry, the fuzzy sphere approach provides access to a wealth of information—such as the renormalization group (RG) flow monotonic $F$-function~\cite{hu2024F}—that is otherwise difficult to obtain through other methods. These developments significantly expand the scope for investigating challenging and intriguing questions in CFTs.

So far, the philosophy of the fuzzy sphere approach has been grounded in the principle of universality, akin to the strategy used in condensed matter models for quantum field theory (QFT). Specifically, to realize a CFT of interest, one begins with an interacting quantum many-body model on the fuzzy sphere that possesses the appropriate global symmetry and ('t~Hooft) anomaly. Then, with a suitable choice of microscopic parameters in the Hamiltonian, the model can flow in the infrared (IR) to the same universality class as the targeted CFT. Despite its successes, gaining more control over the fuzzy sphere approach remains a pressing challenge. In particular, it would be highly beneficial to develop a priori knowledge of how to realize a given QFT/CFT within the fuzzy sphere framework, without relying entirely on numerical methods. An ideal long-term goal is to construct a dictionary that maps all renormalizable Lagrangians to their realizations on the fuzzy sphere. Achieving this would allow the fuzzy sphere approach to move beyond IR universality and address broader QFT questions, such as field theory dualities. As a first step toward this goal, this paper focuses on realizing the free real scalar theory CFT on the fuzzy sphere.

Historically, the idea of regularizing QFTs using non-commutative geometry, originating with Heisenberg, has long been pursued. A major development in this direction is non-commutative field theory~\cite{noncommuQFT,Szabo2003noncomm}, which replaces continuous commutative space with a non-commutative one. This approach stands in contrast to the later-developed lattice regularization, where continuous space is replaced by a discrete lattice. However, non-commutative field theory was found to exhibit a novel feature known as UV-IR mixing, which prevents it from realizing standard QFTs in the IR. In contrast, the recently proposed fuzzy sphere approach successfully realizes the IR of standard QFTs, although its UV lies far from conventional QFT frameworks. This makes it a timely opportunity to revisit fuzzy regularization. The free real scalar theory offers a natural starting point, as it is the simplest QFT that possesses simple yet rich analytical structures. We will show that the fuzzy sphere model can be formulated as an algebra that deforms the harmonic oscillator algebra. Additionally, we provide a simple proof that the fuzzy sphere model is free of UV divergences; that is, both the fields and their correlators remain finite in the continuum (thermodynamic) limit.

The paper is organized as follows. In Sec.~\ref{sec:freeQFT}, we begin with a warm-up discussion of the free scalar CFT on the sphere. In Sec.~\ref{sec:fuzzyfree}, we introduce the fuzzy sphere model for the real scalar CFT fixed point. In particular, we numerically demonstrate that the operator spectrum and correlators of our model agree with the theoretical expectations for the free real scalar CFT. More interestingly, we construct the effective harmonic oscillators of the model and show that the CFT states obey the expected algebra. We further argue that the real scalar in our realization can be understood as the Goldstone mode of spontaneous $U(1)$ symmetry breaking, where the $U(1)$ symmetry becomes the shift symmetry of the real scalar. Notably, unlike typical Goldstone modes, the emergent real scalar in our model is non-compact due to the suppression of vortices. The shift symmetry also forbids all relevant operators, thereby facilitating access to the real scalar CFT fixed point. In Sec.~\ref{sec:fuzzyalgebra}, we introduce the algebra underlying our fuzzy sphere model, demonstrate that it is a deformation of the harmonic oscillator algebra, and show that the model is free of UV divergences. Finally, we propose a wavefunction ansatz motivated by the approximate harmonic oscillator structure, taking the form of a Bogoliubov coherent state. With a slight modification, this wavefunction also exhibits a high overlap with the ground state of the Ising CFT—for example, achieving an overlap greater than $0.99$ for a model with $N=28$ fermions. 

\emph{Note added:} Upon completing this work, we became aware of a parallel study that investigates the free scalar CFT in a different setup~\cite{Taylor_triIsing}.

\section{Warm-up: Free scalar CFT on sphere} \label{sec:freeQFT}

Let us firstly warm up with the free scalar CFT on sphere with the radius $R$, whose Hamiltonian is
\begin{equation} \label{eq:freescalar}
H = \frac{1}{2} \int_{S^2} d^2 \bm x \left(\pi^2(\bm x) - \phi(\bm x) \nabla^2 \phi(\bm x) + M^2 \phi^2(\bm x) \right).
\end{equation}
Here $\pi(\bm x)$ is the canonical momentum, which satisfies the canonical quantization relation,
$[\phi(\bm x_1), \pi(\bm x_2)] = i\delta(\bm x_1 - \bm x_2)$. $M=1/(2R)$ is the mass of the scalar, which vanishes in the infinite radius $R$ limit. This term is called conformal coupling, which gives mass gaps $\Delta E \sim 1/R$ to the excited states, and more importantly, produces correct state-operator correspondence of free scalar CFT~\footnote{The conformal coupling term is marginal ($\Delta=3$), so it is necessary to fine tune it.}. 

One can rewrite the model using harmonic oscillators in the angular momentum space, such that the state-operator correspondence becomes transparent. Firstly, for the quantum fields $\phi(\bm x)$ and $\pi(\bm x)$ we have~\footnote{For the ease of notation, we are using the definition of spherical harmonics without the factor $1/\sqrt{4\pi}$.},
\begin{align} \label{eq:phifield} 
\phi(\bm x) &= \frac{1}{R}  \sum_{\ell=0}^{\infty}\sum_{m=-\ell}^{\ell} \sqrt{\frac{R}{2\ell+1}}(a_{\ell,m} + (-1)^m a^\dagger_{\ell,-m}) Y_{\ell,m}(\theta,\varphi), \\ 
\pi(\bm x ) &= \frac{1}{R}  \sum_{\ell=0}^{\infty}\sum_{m=-\ell}^{\ell} (-i) \sqrt{\frac{2\ell+1}{4R}}(a_{\ell,m} - (-1)^m a^\dagger_{\ell,-m}) Y_{\ell,m}(\theta,\varphi). \label{eq:pifield}
\end{align}
here $a_{\ell,m}$ and $a^\dagger_{\ell,m}$ are the familiar harmonic oscillators,
\begin{equation}
[a_{\ell_1,m_1}, a^\dagger_{\ell_2,m_2}] = \delta_{\ell_1,\ell_2} \delta_{m_1,m_2}.
\end{equation}
It is straightforward to show that, $\phi(x)$ and $\pi(x)$ satisfy the canonical quantization relation,
\begin{equation}
[\phi(\bm x_1), \pi(\bm x_2)] = i\frac{1}{R^2} \delta(\varphi_1 -\varphi_2)\delta(\cos\theta_1 - \cos\theta_2) = i\delta(\bm x_1 - \bm x_2) .
\end{equation}
Furthermore, the Hamiltonian becomes
\begin{equation}
H = \sum_{\ell=0}^\infty \sum_{m=-\ell}^\ell \left( \frac{\ell+1/2}{R} a^\dagger_{\ell,m} a_{\ell,m} + \frac{\ell+1/2}{2R} [a_{\ell,m}, a^\dagger_{\ell,m}] \right). 
\end{equation}
The last term is the usual vacuum Casimir energy, which we remove as is standard practice in QFT textbooks. The ground state $|\bm{0}\rangle$ of the Hamiltonian is naturally the vacuum of the harmonic oscillators, satisfying $a_{\ell, m} |\bm{0}\rangle = 0$. The excited states are obtained by acting with creation operators on the vacuum. We are now ready to examine the state-operator correspondence of the free scalar CFT, namely that the energy of each CFT state is related to the scaling dimension of the corresponding operator via the relation $E = \Delta / R$. Below are some examples,
\begin{enumerate}
    \item Conformal family of $\phi(\bm x)$ ($\Delta=1/2$). The primary state is $|\phi\rangle = a^\dagger_{0,0} |\bm 0\rangle$, and its energy is $1/(2R)$. Its descendant $\partial_{\mu_1}\cdots \partial_{\mu_\ell} \phi(\bm x)$ corresponds to the state $a^\dagger_{\ell,m} |\bm 0\rangle$, whose energy is $(\ell+1/2)/R$. 
    \item Conformal family of $\phi^2(\bm x)$ ($\Delta=1$). The primary state is $|\phi^2\rangle = (a^\dagger_{00})^2/2 |\bm 0\rangle$, and its energy is $1/R$. The descendant $\partial_{\mu_1}\cdots \partial_{\mu_\ell} \phi^2(\bm x)$ is $a^\dagger_{\ell,m} a^\dagger_{0,0} |\bm 0\rangle$, and its energy is $(\ell+1)/R$. Other descendants can be written in a similar yet more complicated way.
    \item The primary state of $\phi^n(\bm x)$ ($\Delta=n/2)$ is $|\phi^n\rangle =(a^\dagger_{00})^n/n! |\bm 0\rangle $ and its energy is $n/(2R)$.
\end{enumerate}
We can also directly compute correlators, for example,
\begin{equation}
\langle \phi(\bm x_1) \phi(\bm x_2) \rangle =\frac{1}{R} \sum_{\ell,m} \frac{1}{2\ell+1} \bar Y_{\ell,m}(\theta_1,\varphi_1) Y_{\ell,m}(\theta_2, \varphi_2) = \frac{1}{2R\sin(\gamma_{12}/2)},
\end{equation}
where $\gamma_{12}$ is the angle between $\bm x_1$ and $\bm x_2$. The two-point correlator of the canonical momentum $\pi(\bm x)$ is,
\begin{equation}\label{eq:picorr}
    \langle \pi(\bm x_1) \pi(\bm x_2)\rangle = \frac{1}{4R^3} \sum_{\ell,m} (2\ell+1) \bar Y_{\ell,m}(\theta_1,\varphi_1) Y_{\ell,m}(\theta_2, \varphi_2),
\end{equation}
which is divergent. One way to see the divergence of the series summation is to examine the correlator between the north and south pole,
\begin{equation}
  \langle \pi(\bm x_1=\textrm{North pole}) \, \pi(\bm x_2=\textrm{South pole})\rangle   =\frac{1}{4R^3} \sum_{\ell} (-1)^\ell (2\ell+1)^2. 
\end{equation}  
Nevertheless, from the basics of CFT we know $\pi(\bm x)=i\partial_\tau \phi(\bm x)$ is the descendant of $\phi(\bm x)$, whose correlator is $\langle \pi(\bm x_1) \pi(\bm x_2)\rangle = -\frac{1}{(2R\sin(\gamma_{12}/2))^3}$. One way to get the correct correlator from Eq.~\eqref{eq:picorr} is to regularize it with a regularization factor $K_\ell=\sqrt{2s+1} \tj{s}{s}{\ell}{s}{-s}{0} $~\footnote{Here $\tj{s}{s}{\ell}{s}{-s}{0}$ is the Wigner-3j symbol.}, 
\begin{align}\label{eq:regpicorr}
\langle \pi(\bm x_1) \pi(\bm x_2)\rangle &= \frac{1}{4R^3} \lim_{s\rightarrow\infty}\sum_{\ell=0}^{2s}\sum_{m=-\ell}^\ell (2\ell+1) K_\ell \bar Y_{\ell,m}(\theta_1,\varphi_1) Y_{\ell,m}(\theta_2, \varphi_2) \nonumber \\  &=  -\frac{1}{(2R\sin(\gamma_{12}/2))^3}.   
\end{align}
This regularization scheme is using the property that 
\begin{align}
&\lim_{s \rightarrow \infty} K_\ell  \approx e^{-\ell^2/s},
\end{align}
which is $1$ for small $\ell $ ($<\sqrt{s}$), but vanishes exponentially fast at large $\ell$ ($>\sqrt{s}$). 
As we will discuss later, this regularization factor emerges naturally within the fuzzy sphere regularization, potentially shedding light on its underlying mechanism.

At last, we note that without the conformal coupling term ($M=0$), the free scalar theory has a symmetry called shift symmetry, $\phi\rightarrow \phi+c$, which is a $\mathbb{R}$-symmetry. The conserved current of the shift symmetry is $\partial_\mu \phi$, and the conserved charge is $\int_{S^2} d^2\bm x \, i\partial_\tau \phi(\bm x) = \int_{S^2}  d^2\bm x  \, \pi(\bm x) $. This shift symmetry will play a key role in our realization of the free scalar theory on the fuzzy sphere.

\section{Free scalar CFT on the fuzzy sphere} \label{sec:fuzzyfree}
\subsection{Hamiltonian and symmetry}
As in the case of the 3D Ising CFT on the fuzzy sphere~\cite{ZHHHH2022}, we consider a model on the sphere with fermions $\Psi^\dagger(\bm{x}) = (\psi^\dagger_{\uparrow}(\bm{x}), \psi^\dagger_{\downarrow}(\bm{x}))$ that carry an isospin degree of freedom and are subject to a magnetic monopole with flux $4\pi s$. The magnetic monopole creates Landau levels for the fermions, and we focus on the lowest Landau level (LLL), which consists of $N = 2s + 1$ orbitals forming a spin-$s$ representation of the $SO(3)$ sphere rotation symmetry~\cite{Sphere_LL_Haldane}. We refer the reader to Ref.~\cite{ZHHHH2022} for more detailed descriptions of the fuzzy sphere regularization. In this paper, we present the formulation in explicit terms—that is, we define the model in terms of its Hilbert space and operators.

First of all, we have fermion operators $c_{m,\sigma}$, $m=-s,-s+1,\cdots, s$, $\sigma=\uparrow,\downarrow$. These fermion operators follow the standard anticommutation relations of fermions, $\{ c_{m_1,\sigma_1}, c_{m_2,\sigma_2} \} = \delta_{m_1,m_2}\delta_{\sigma_1,\sigma_2}$. Importantly, they form a spin-$s$ representation of $SO(3)$ sphere rotation, corresponding to the $2s+1$ states on the LLL. We consider the case that has $N=2s+1$ fermions filled in the LLL, so the many-body Hilbert space is spanned by $\prod_{i=1}^{N} c_{m_i,\sigma_i}^\dag |0\rangle$. 

The fermion field $\psi_\sigma(\bm x)$ in the real space is defined as
\begin{equation}
    \psi_\sigma(\bm x) =\frac{1}{\sqrt{2s+1}}\sum_{m=-s}^s c_{m,\sigma} Y^{(s)}_{s,m}(\bm x),
    \end{equation}
$Y^{(s)}_{s,m}(\bm x)$ is the monopole harmonics (omitting the $1/\sqrt{4\pi}$ factor),
 \begin{equation}\label{eq:monopoleHarmonics}
Y^{(s)}_{s,m}(\bm x) = \sqrt{\frac{(2s+1)!}{(s+m)!(s-m)!}} e^{im\varphi} \cos^{s+m}\left(\frac{\theta}{2}\right)\sin^{s-m}\left(\frac{\theta}{2}\right)
 \end{equation}
 which is the wave-function of LLL. Here and throughout, we consider the field $\psi_\sigma(\bm{x})$ at zero time, $\tau = 0$, unless otherwise specified.  It is straightforward to verify that the fermion field in real space does not satisfy the usual anti-commutation relations of fermions anymore.
 
We note that, in this formulation, the field has a well-defined real space coordinate, in contrast to the usual expectation that coordinates become fuzzy due to the uncertainty principle in non-commutative geometry. Here, the non-commutative nature is instead manifested through non-locality: fields at different positions do not anti-commute, i.e., $\{ \psi^\dag_{\sigma}(\bm{x}), \psi_{\sigma}(\bm{y}) \} \neq 0$ for $\bm{x} \neq \bm{y}$ when $s$ is finite. Nevertheless, in the continuum limit, i.e., as $s \rightarrow \infty$, the fields at different positions begin to anti-commute once again, thereby restoring locality consistent with the familiar commutative geometry.

Next we can use the fermion field $\Psi^\dagger(\bm x) = (\psi^\dagger_{\uparrow}(\bm x), \psi^\dagger_{\downarrow}(\bm x))$ to construct composite field, especially density field $n^0(\bm x) = \Psi^\dagger(\bm x) \Psi(\bm x) $ and $n^\alpha(\bm x) =\Psi^\dagger(\bm x) \sigma^\alpha \Psi(\bm x) $ with $\sigma^{\alpha=x,y,z}$ be the Pauli matrices. Using these density fields, we can write down the Hamiltonian that realizes the free scalar CFT,
\begin{equation}\label{eq:fuzzy_gaussiantext}
H = \int_{S^2} d^2 \bm x \left( (n^0(\bm x))^2 + U n^y(\bm x) \nabla^2 n^y(\bm x) + h \frac{n^z(\bm x)}{R^2}   \right).
\end{equation}
 $U$ and $h$ are (order-1) free parameters we can tune to hit the free scalar CFT fixed point. Here $R$ is the radius of sphere, and we will simply replace $R$ with $\sqrt{2s+1}$ when switching to the fermion $c_{m,\sigma}$ representations. The continuum limit corresponds to $s\rightarrow\infty$. 
 
 The Ising CFT can be realized by the Hamiltonian~\footnote{The original paper~\cite{ZHHHH2022} includes an additional term, $U'\, n^0(\bm x) \nabla^2 n^0(\bm x)$, which corresponds to a deformation by an irrelevant operator. In practice, the Hamiltonian with or without this term yields high-quality results for the Ising CFT.}, 
 \begin{equation}\label{eq:fuzzy_Ising}
H = \int_{S^2} d^2 \bm x \left( (n^0(\bm x))^2 - U' n^x(\bm x) \nabla^2 n^x(\bm x) + h' n^z(\bm x)   \right).
\end{equation}
It is interesting that these two Hamiltonians are structurally very similar, differing in three key aspects: (1) $n^x$ is replaced by $n^y$; (2) the signs of the spin interaction terms are opposite; and (3) the coupling strength of the transverse field $n^z$ differs in scale.

In the fermion representations, it is easy to figure out the symmetries of our model,
\begin{enumerate}
    \item $SO(3)$ sphere rotation: $c_{m,\sigma}$, $m=-s,-s+1,\cdots,s$ form a spin-$s$ represention of $SO(3)$. 
    \item Ising $\mathbb{Z}_2$ symmetry: $(c_{m,\uparrow}, c_{m,\downarrow}) \rightarrow (c_{m,\uparrow}, -c_{m,\downarrow}) =(c_{m,\uparrow}, -c_{m,\downarrow}) \sigma^z$.
    \item Time-reversal symmetry~\footnote{It is not the canonical time-reversal symmetry, which is broken by the magnetic field of monopole in our original problem. This new time-reversal is nothing but that our Hamiltonian only contains real numbers.}: complex conjugate,  $i\rightarrow -i$, and it keeps all the fermion operators $c_{m,\sigma}$ intact. 
		\item Particle-hole symmetry (unitary): $(c_{m,\uparrow}, c_{m,\downarrow}) \rightarrow (c_{m,\downarrow}^\dag, -c_{m,\uparrow}^\dag)$. It acts as an improper $\mathbb{Z}_2$ of the sphere rotation $O(3)$, and it can be identified as the space parity of the 3D CFT.
\end{enumerate}
We can easily identify symmetry quantum numbers of the four density fields,
\begin{enumerate}
    \item $n^z(\bm x)$: even under the Ising $\mathbb Z_2$, time-reversal and particle-hole symmetry.
     \item $n^x(\bm x)$: odd under the Ising $\mathbb Z_2$, but even under the  time-reversal and particle-hole symmetry.   
    \item $n^y(\bm x)$: odd under the Ising $\mathbb Z_2$ and  time-reversal, but even under the particle-hole symmetry.
    \item $n^0(\bm x)$: odd under the particle-hole (after subtracting a constant background density), but even under the Ising $\mathbb Z_2$ and  time-reversal symmetry.
\end{enumerate}
So from the symmetry point of view, the density field $n^{x,y,z}(\bm x)$ corresponds to familiar fields in the real scalar theory~\footnote{The operator $n^0(\bm{x})$ could either correspond to parity-odd operators with a relatively high scaling dimension, or it may decouple in the IR, in which case its correlators decay exponentially. At present, we have not determined which of these scenarios is correct.},
 \begin{equation}
n^x(\bm x) \sim \phi(\bm x)+\cdots, \quad\quad n^y(\bm x) \sim \pi(\bm x)+\cdots, \quad\quad n^z(\bm x) \sim I+ \lambda \phi^2(\bm x)+\cdots,
 \end{equation}
 where $\cdots$ stands for the fields with the same quantum numbers. Therefore, the last term $n^z(\bm{x})/R^2$ in our fuzzy sphere Hamiltonian can be interpreted as the conformal coupling term, analogous to $\phi^2(\bm{x})/R^2$ in the real scalar theory. If the conformal coupling term vanishes ($h=0$), and the Hamiltonian acquires an additional $U(1)$ rotational symmetry, whose associated charge is given by
\begin{equation}
\int_{S^2} n^y(\bm{x}) = \sum_{m=-s}^{s} (c_{m,\uparrow}^\dag, c_{m,\downarrow}^\dag)\, \sigma^y \begin{pmatrix} c_{m,\uparrow} \\ c_{m,\downarrow} \end{pmatrix}.
\end{equation}
As we will discuss later in Sec.~\ref{sec:shiftsymmetry}, this $U(1)$ symmetry in the UV becomes an $\mathbb{R}$-symmetry (shift symmetry) in the IR, analogous to the Goldstone phase in which a $U(1)$ symmetry is spontaneously broken.\footnote{We thank Chong Wang for the discussion on this.} This is reflected in the correspondence $n^y(\bm{x}) \sim \pi(\bm{x}) + \cdots$, where the charge associated with the shift symmetry in the real scalar theory is $\int_{S^2} \pi(\bm{x})$.

 At last, we remark that from the perspective of symmetry, the first term $(n^0(\bm x))^2$ in the Hamiltonian seems to be unnecessary. But numerically we find it is important to include $(n^0(\bm x))^2$ in order to suppress non-CFT states. These non-CFT states typically have very large $SO(3)$ Lorentz spin, and are likely to be the magneton-rotons in quantum Hall systems~\cite{GMP}.

\subsection{Operator spectrum}

\begin{figure}
    \centering
    \includegraphics[width=0.5\linewidth]{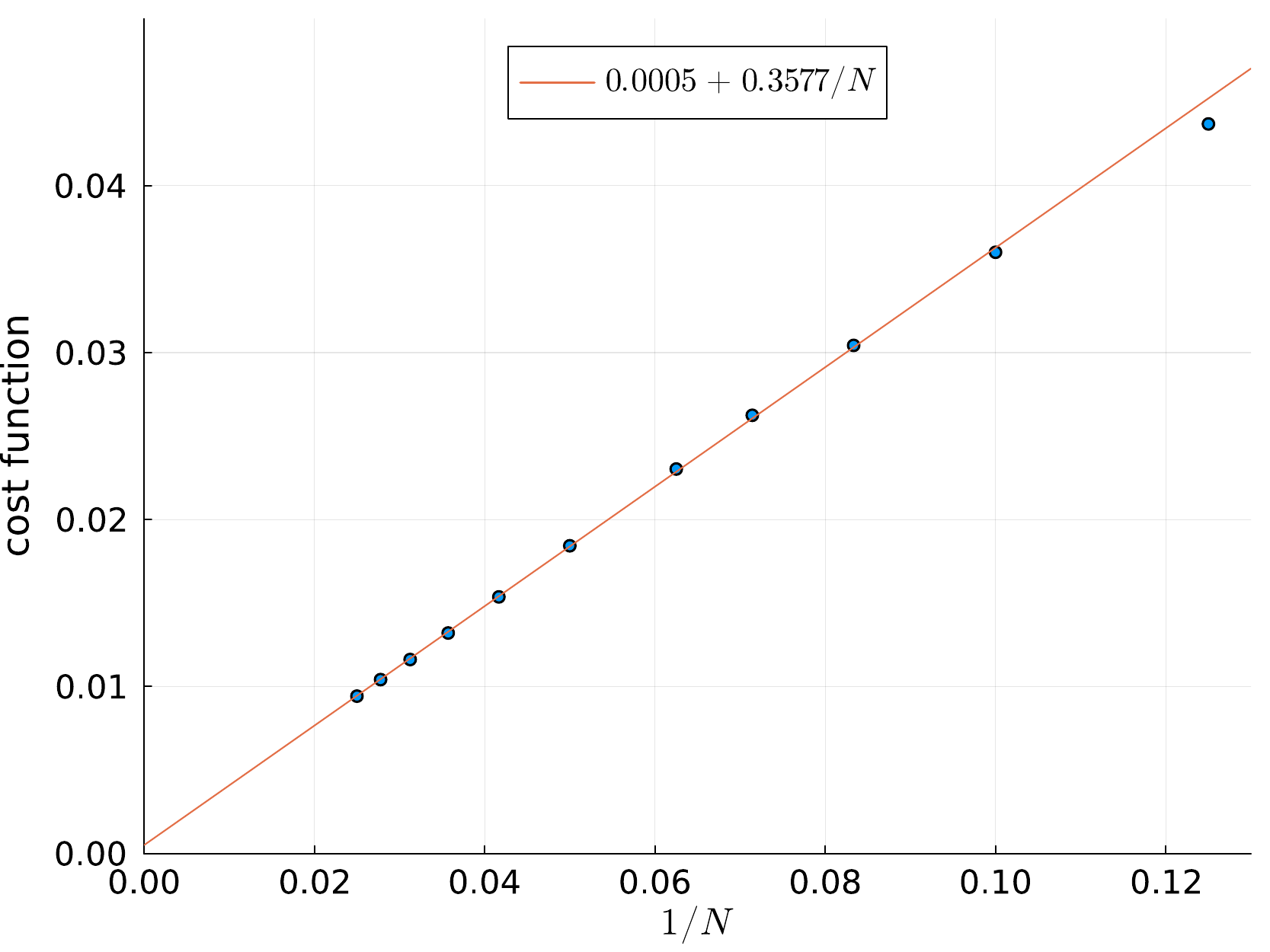}
    \caption{\label{fig:cost} A cost function characterizing the dependency between the free real scalar CFT and  the spectrum of the fuzzy sphere model Eq.~\eqref{eq:fuzzy_gaussiantext} at $(U, h)=(0.18, 0.42)$. The cost function behaves almost linearly with $1/N$, and a linear extrapolation gives nearly $0$ at infinite $N$.  }
\end{figure}

In our fuzzy sphere Hamiltonian, Eq.~\eqref{eq:fuzzy_gaussiantext}, when the conformal coupling term is omitted, the $U(1)$ symmetry emerges as a shift symmetry, which forbids operators such as $\phi(\bm{x})^2$, $\phi(\bm{x})^4$, and other higher-order terms.
 However, operators like $\phi(\bm{x})^2/R^2$ and $\phi(\bm{x})^4/R^2$ are allowed by symmetry due to the explicit $U(1)$ breaking introduced by the term $n^z(\bm{x})/R^2$. As a result, no symmetry-allowed relevant operators exist, but a marginal operator of the form $\phi^2/R^2$ remains, which must be fine-tuned. In practice, we find that tuning two parameters—$U$ and $h$ in Eq.~\eqref{eq:fuzzy_gaussiantext}—significantly improves the quality of the conformal spectrum. To identify the optimal values of $U$ and $h$, we minimize a cost function defined by the deviation of our spectrum from that of the free real scalar CFT. Specifically, we use the energies of seven low-lying states, corresponding to the operators $\phi$, $\partial_\mu \phi$, $\partial_\mu \partial_\nu \phi$, $\phi^2$, $\partial_\mu \phi^2$, $\partial_\mu \partial_\nu \phi^2$, and $T_{\mu\nu}$, to define the cost function (see Appendix~\ref{sec:cost} for its precise definition). We  compute the low-lying states of finite system sizes $N=2s+1\sim R^2$, using both exact diagonalization (for $N \leq 16$) and DMRG~\cite{SWhite1992} (for $N > 16$) \footnote{We use a bond dimension of $8000$, which yields well-converged results for both ground and excited states.} facilitated by the numerical package FuzzifiED~\cite{zhou2025fuzzifiedjuliapackage} and ITensor~\cite{itensor}.  Based on a brute-force optimization of small sizes ($N\le 16$), we choose $(U, h) = (0.18, 0.42)$ for presenting the numerical results. The cost function at this point clearly scales to $0$ at infinite $N=2s+1$, as shown in Fig.~\ref{fig:cost}.

\begin{figure}
    \centering
    \includegraphics[width=0.49\linewidth]{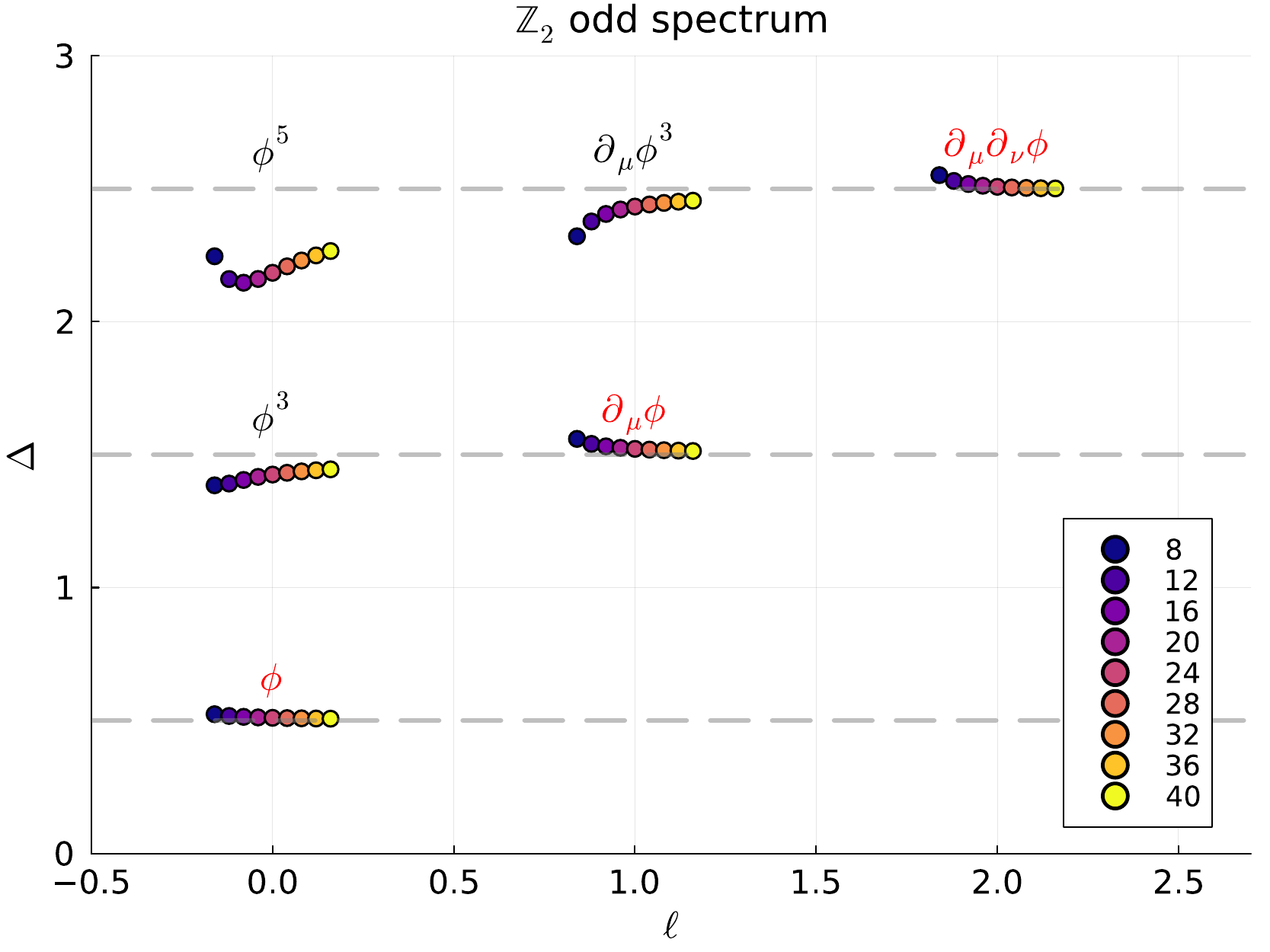}
    \includegraphics[width=0.49\linewidth]{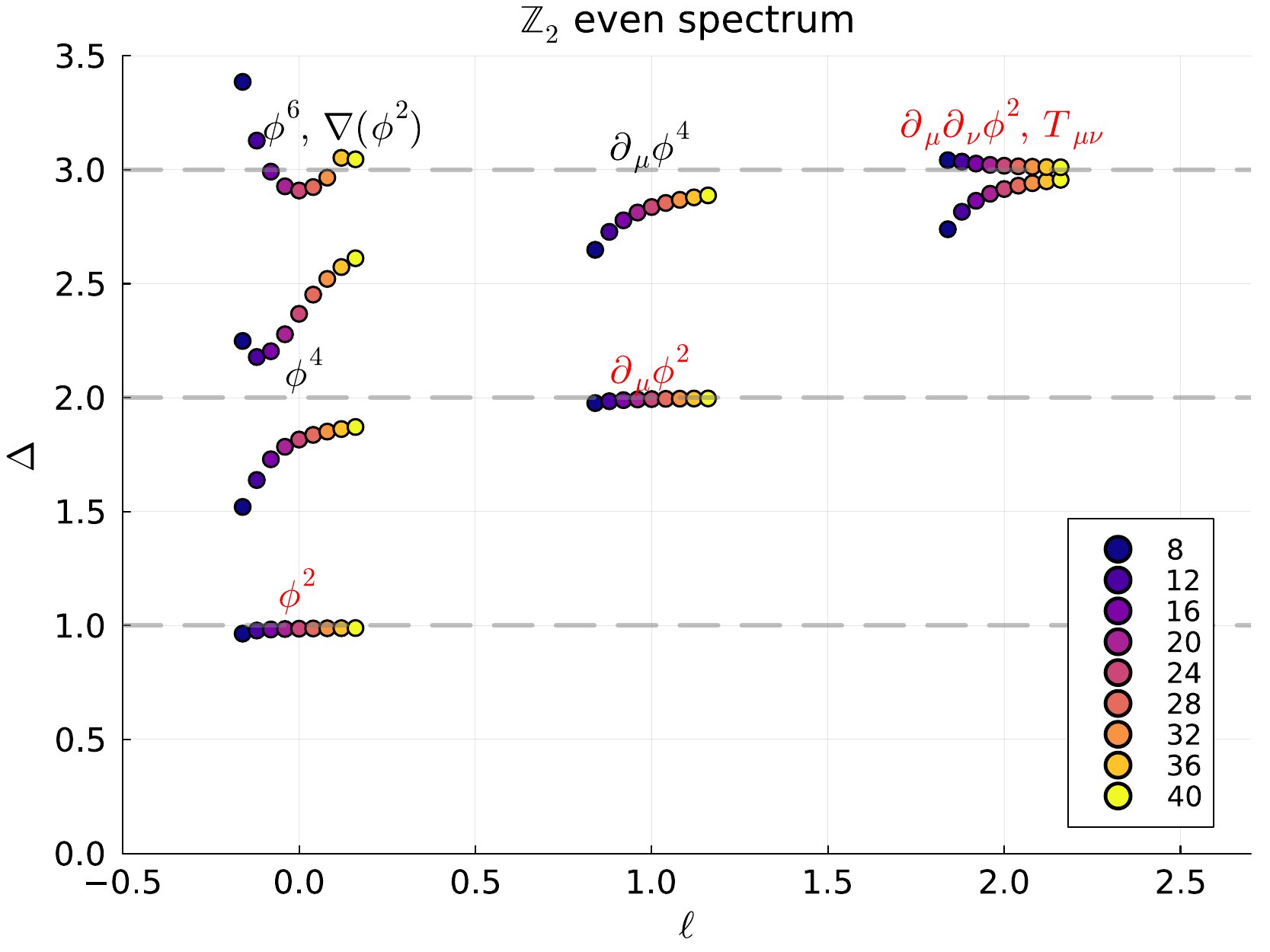}
    \caption{Low lying operator spectrum in the $\mathbb Z_2$ odd and even sectors at different system sizes $N=8, 12, \cdots, 40$. }
    \label{fig:spectrum}
\end{figure}

We now turn to the low-lying operator spectrum. Fig.~\ref{fig:spectrum} shows the low-lying operators with scaling dimensions $\Delta \leq 3$ for system sizes in the range $8 \leq N \leq 40$. The states highlighted in red in the figure are those used for the parameter optimization, and their energies are remarkably close to the theoretical values, even at relatively small system sizes.\footnote{One state at $\Delta = 3$, $\ell = 2$ shows a noticeably larger discrepancy and is the dominant contributor to the cost function.} Although the parameters were chosen specifically to optimize this set of states, it is notable that only two parameters were fine-tuned while achieving good agreement across seven distinct operators. It is also worth noting that the optimization was carried out only for small system sizes ($N \leq 16$), yet the agreement with the theoretical spectrum continues to improve for larger system sizes in the range $20 \leq N \leq 40$.
 More importantly, other states—not used in the optimization—also converge toward the expected theoretical values as the system size increases, despite exhibiting larger finite-size effects. We therefore conclude that the spectrum of our fuzzy sphere model closely matches that of the free scalar CFT, although the finite-size effects are somewhat more pronounced than in the previously studied 3D Ising CFT. Further improvement of the spectrum could be achieved by fine-tuning additional parameters, which we leave for future investigation. It is also interesting to use conformal perturbation to understand the discrepancy at the finite system sizes~\cite{Lauchli_CPT}. 

\subsection{Operators and correlators} \label{sec:correlators}

To confirm the emergence of the free real scalar CFT in our fuzzy sphere model, it is also important to study operators and their correlators. In general, each UV operator can be expressed as a superposition of IR operators with the same quantum numbers,
\begin{equation}
O_{\text{UV}}(\bm{x}) = \sum_{i} \alpha_i\, O^i_{\text{IR}}(\bm{x}).
\end{equation}
When computing the correlators of a UV operator, all IR components contribute with a weight scaling as $N^{-\Delta/2}$, where $\Delta$ is the scaling dimension. Therefore, in the continuum limit $N = 2s + 1 \rightarrow \infty$, the IR operator with the smallest scaling dimension dominates the correlator. Consequently, we have
\begin{align} 
\langle n^x(\bm{x}_1)\, n^x(\bm{x}_2) \rangle &= \frac{\alpha_1}{2\sqrt{2s+1}\, \sin(\gamma_{12}/2)} \sim \langle \phi(\bm{x}_1)\, \phi(\bm{x}_2) \rangle, \\
\langle n^z(\bm{x}_1)\, n^z(\bm{x}_2) \rangle - \langle n^z(\bm{x}_1) \rangle^2 &= \frac{\alpha_2}{\left(2\sqrt{2s+1}\, \sin(\gamma_{12}/2)\right)^2} \sim \langle \phi^2(\bm{x}_1)\, \phi^2(\bm{x}_2) \rangle, \\
\langle n^y(\bm{x}_1)\, n^y(\bm{x}_2) \rangle &= -\frac{\alpha_3}{\left(2\sqrt{2s+1}\, \sin(\gamma_{12}/2)\right)^3} \sim \langle \pi(\bm{x}_1)\, \pi(\bm{x}_2) \rangle,
\end{align}
where $\alpha_{1,2,3}$ are order-one positive numerical constants, and $\gamma_{12}$ is the relative angle between $\bm x_1$ and $\bm x_2$.  

\begin{figure}
    \centering
    \includegraphics[width=0.24\linewidth]{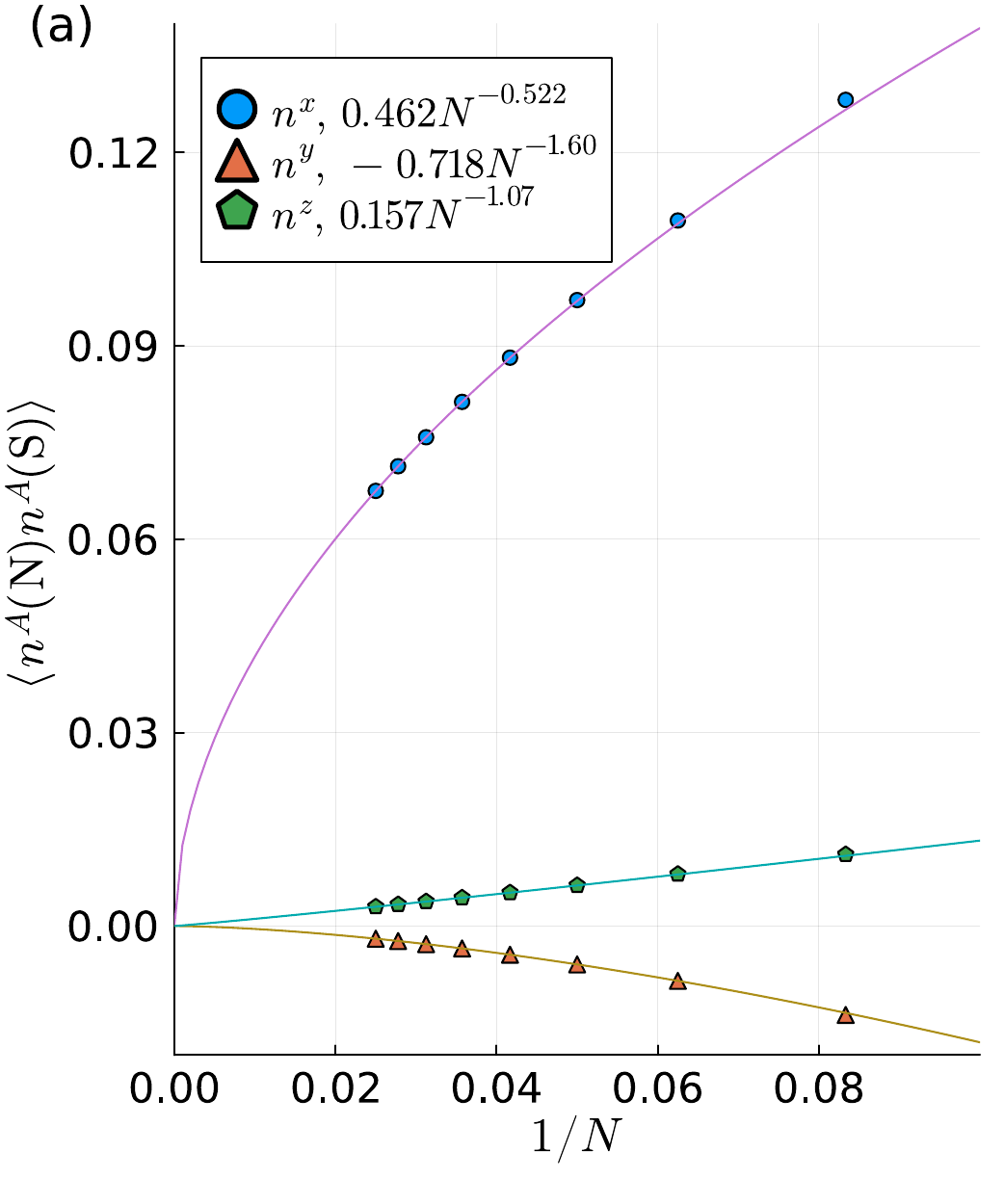}
    \includegraphics[width=0.24\linewidth]{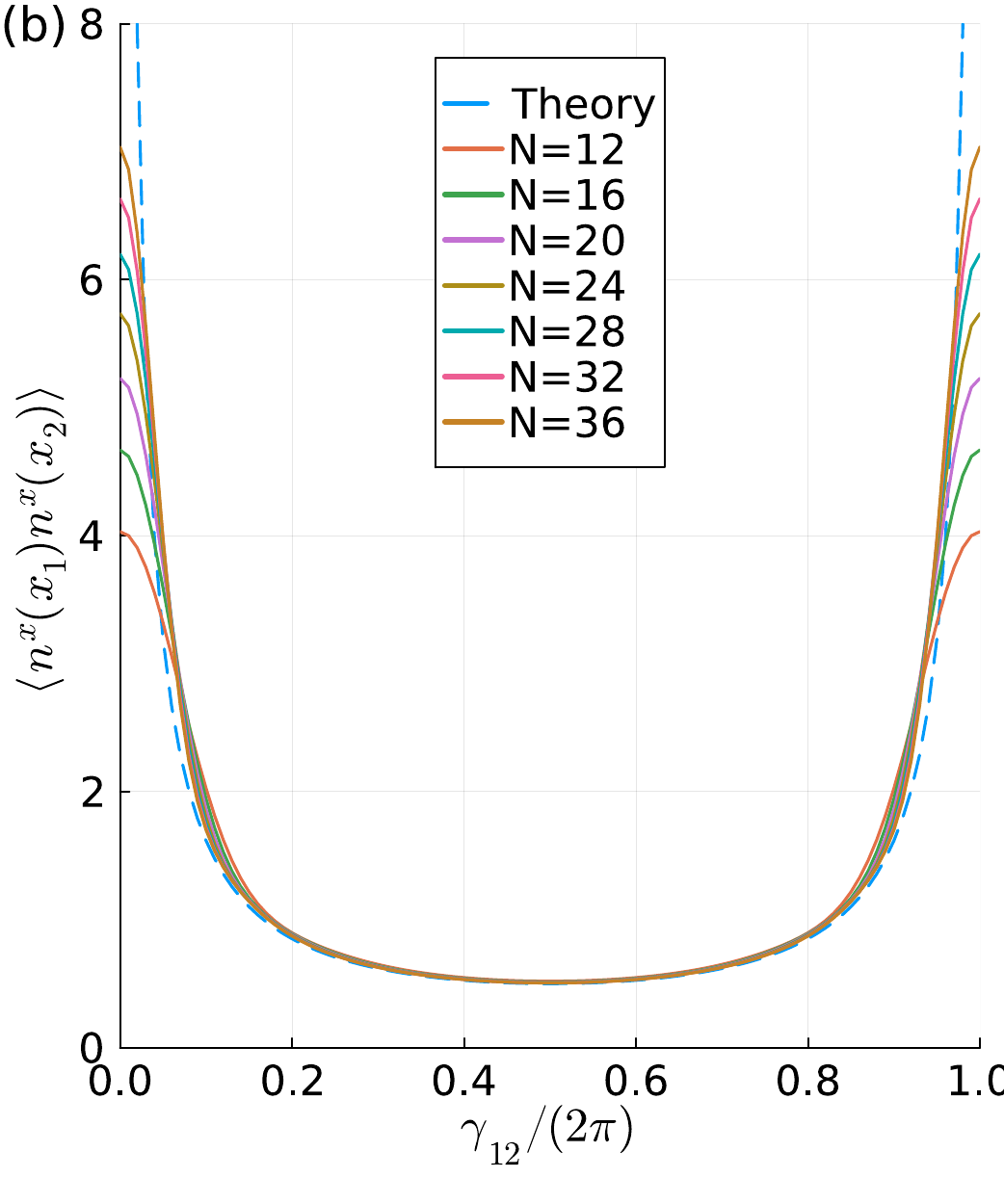}
    \includegraphics[width=0.24\linewidth]{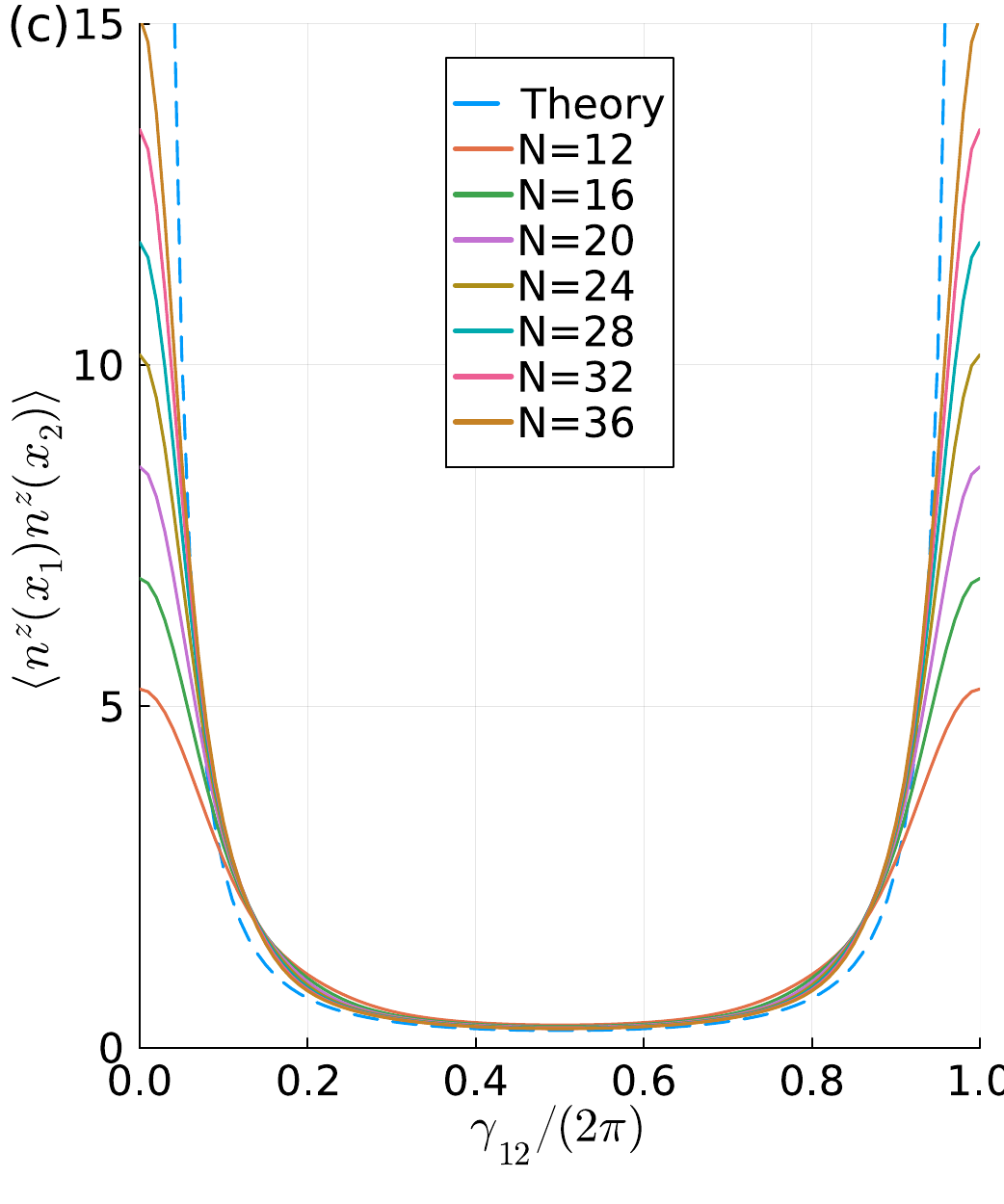}
        \includegraphics[width=0.24\linewidth]{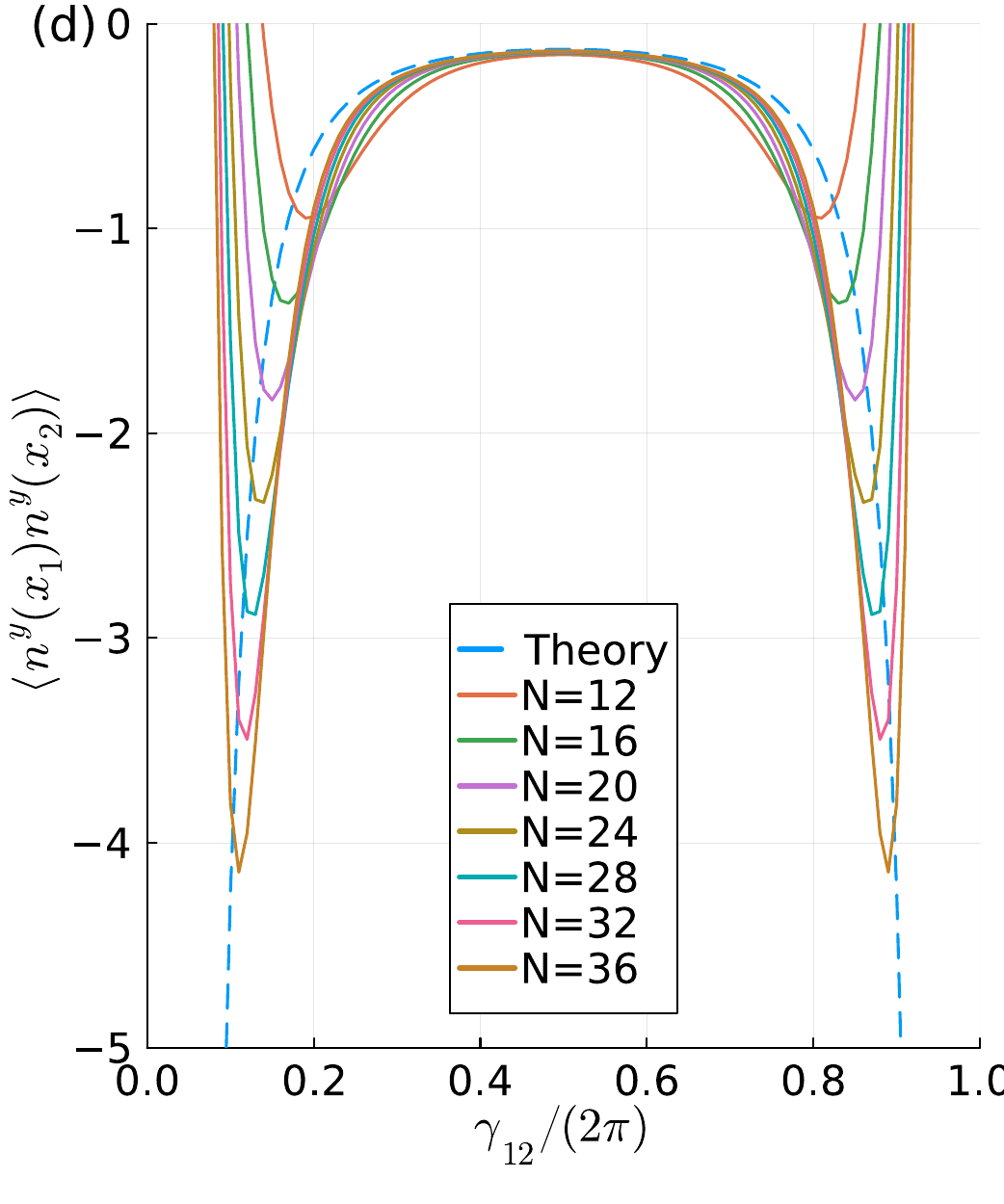}
    \caption{(a) The size dependence of antipodal correlators. (b)-(d) Dimensionless two-point correlators of $n^x$, $n^y$ and $n^z$. }
    \label{fig:correlator}
\end{figure}

These correlators exhibit nontrivial dependence on both the system size $N = 2s + 1$ and the relative angle $\gamma_{12}$. Fig.~\ref{fig:correlator}(a) shows the system-size dependence of the antipodal two-point correlators of $n^{x}$, $n^{y}$, and $n^{z}$, which are consistent with the theoretical expectations, despite small discrepancies.
Next, we evaluate the dimensionless two-point correlators introduced in Ref.~\cite{Han2023Conformal}, defined as:
\begin{align}
\frac{\langle n^x(\bm{x}_1)\, n^x(\bm{x}_2) \rangle}{|\langle \phi |n^x(\bm{x}_1)| \bm{0} \rangle|^2} &= \frac{1}{2 \sin(\gamma_{12}/2)}, \\
\frac{\langle n^z(\bm{x}_1)\, n^z(\bm{x}_2) \rangle - \langle n^z(\bm{x}_1) \rangle^2}{|\langle \phi^2 |n^z(\bm{x}_1)| \bm{0} \rangle|^2} &= \frac{1}{4 \sin^2(\gamma_{12}/2)}, \\
\frac{\langle n^y(\bm{x}_1)\, n^y(\bm{x}_2) \rangle}{|\langle \phi |n^y(\bm{x}_1)| \bm{0} \rangle|^2} &= -\frac{1}{8 \sin^3(\gamma_{12}/2)}.
\end{align}
Fig.~\ref{fig:correlator}(b)-(d) presents these dimensionless two-point correlators, showing excellent agreement with the theoretical predictions for the free real scalar theory, especially as $\theta \to \pi$. These results further support the conclusion that our fuzzy sphere model successfully realizes the free real scalar CFT.

It is also interesting to quantitatively understand how the correlator converges to that of the free scalar theory as $s$ increases. One natural way to do this is to compare the correlators via their spherical harmonic expansions,
\begin{equation} \label{eq:haromincexpansion}
\langle O(\bm{x}_1)\, O(\bm{x}_2)\rangle = \sum_{\ell = 0}^{\infty} \sum_{m = -\ell}^{\ell} A_\ell\, \bar{Y}_{\ell, m}(\bm{x}_1)\, Y_{\ell, m}(\bm{x}_2).
\end{equation}
For the free scalar correlator $\langle \phi(\bm{x}_1)\, \phi(\bm{x}_2) \rangle = 1/(2\sin \gamma_{12})$, we have $A_\ell = 1/(2\ell + 1)$. In our fuzzy sphere model, the operator $n^x(\bm{x})$ corresponds to $\phi(\bm{x})$, and its dimensionless two-point correlator takes the form
\begin{equation}\label{eq:numericalharomincexpansion}
\frac{\langle n^x(\bm{x}_1)\, n^x(\bm{x}_2) \rangle}{\langle \phi | n^x(\bm{x}) | 0 \rangle^2} = \sum_{\ell = 0}^{2s} \sum_{m = -\ell}^{\ell} B_\ell\, \bar{Y}_{\ell, m}(\bm{x}_1)\, Y_{\ell, m}(\bm{x}_2).
\end{equation} 
In the limit $s \to \infty$, the $n^x(\bm{x})$ correlator is expected to converge to that of $\phi(\bm{x})$, implying $\lim_{s \to \infty} B_\ell = 1/(2\ell + 1)$.

\begin{figure}
    \centering
    \includegraphics[width=0.495\linewidth]{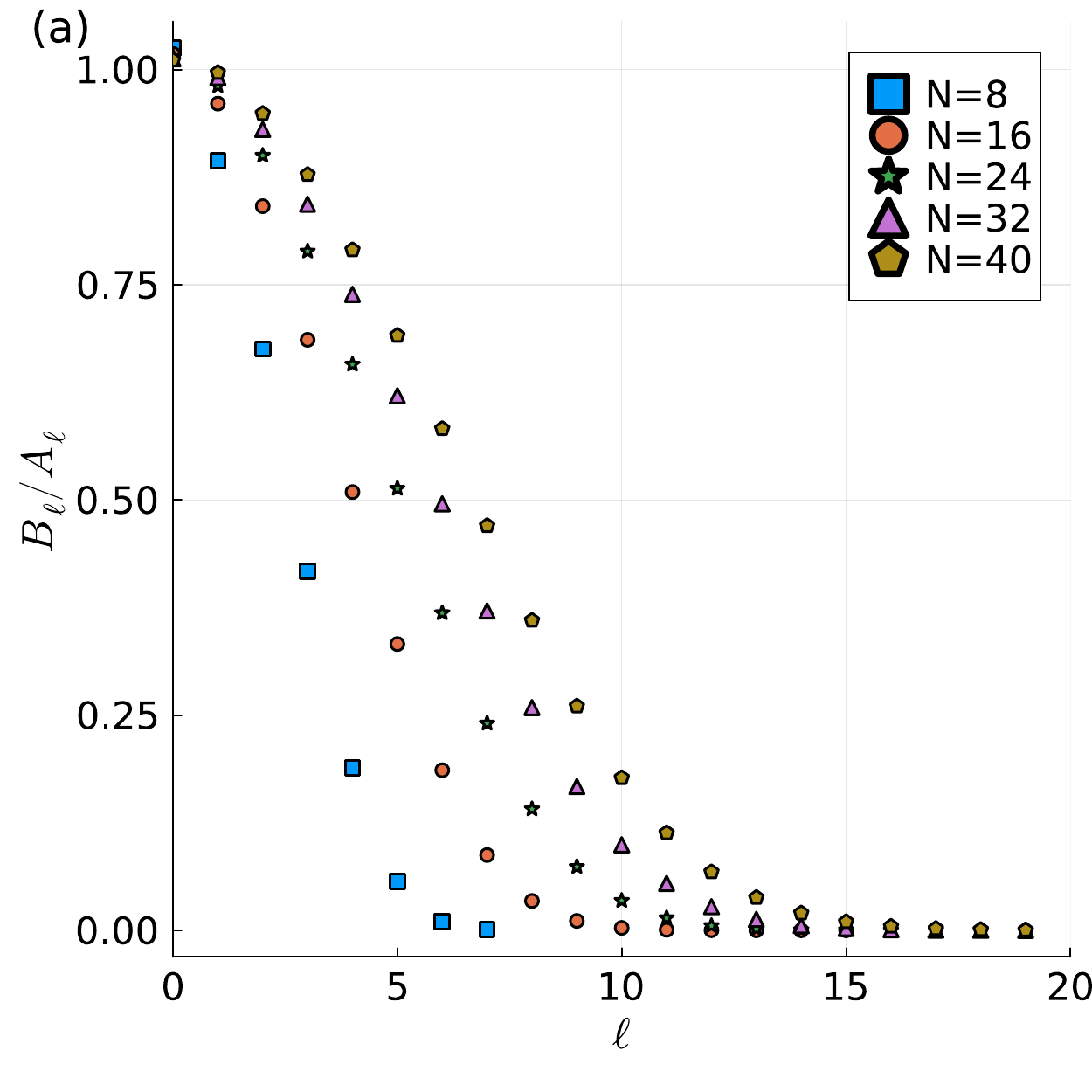}
        \includegraphics[width=0.495\linewidth]{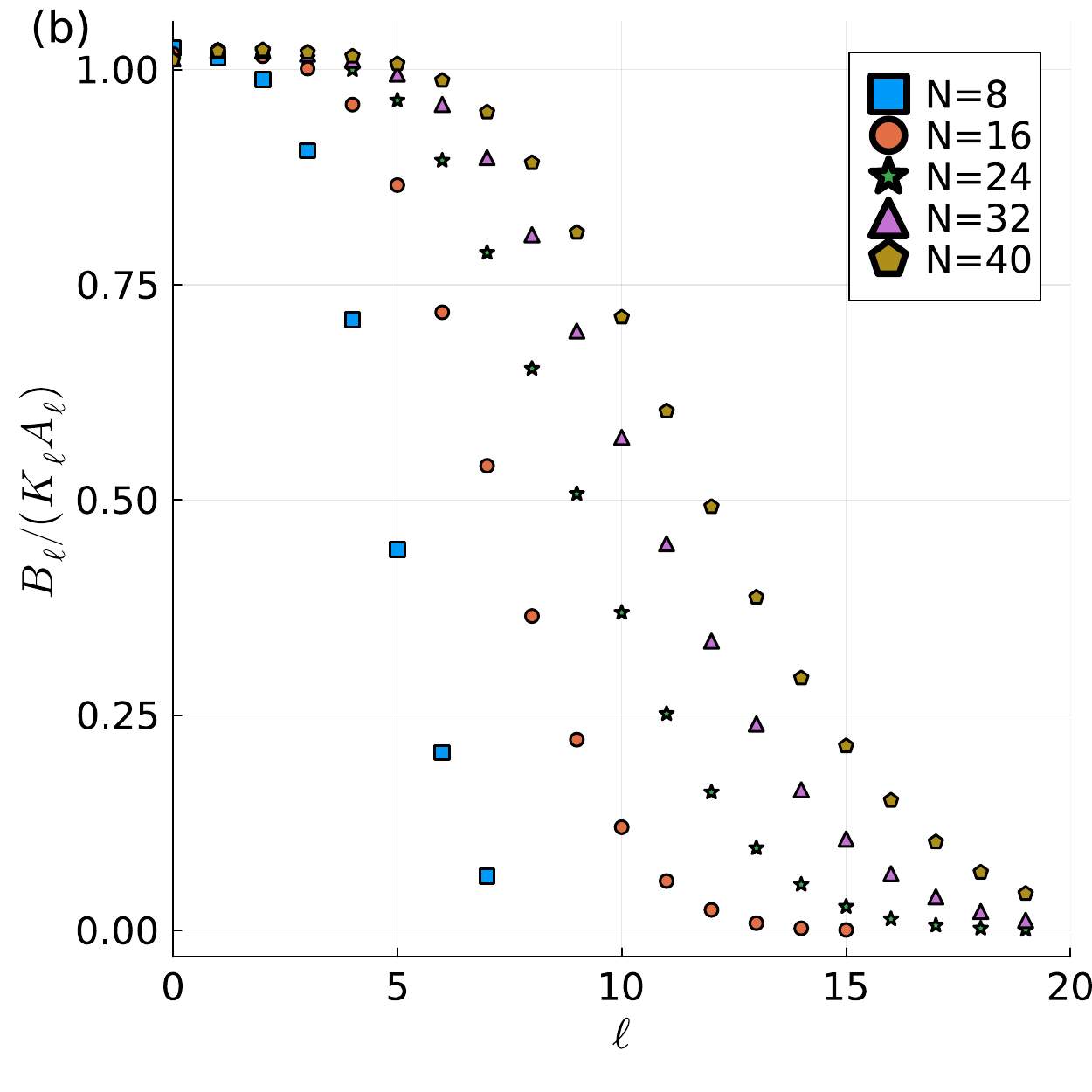}
    \caption{We compare the coefficients in the spherical harmonic expansion of the two-point correlator as given in Eq.~\eqref{eq:haromincexpansion}, Eq.~\eqref{eq:numericalharomincexpansion}, and Eq.~\eqref{eq:regularizedharmonicexpansion}. The numerical value $B_\ell$ converges to the regularized version $K_\ell A_\ell$.}
    \label{fig:coefficients}
\end{figure}

Fig.~\ref{fig:coefficients}(a) shows the ratio $B_\ell / A_\ell$ for different system sizes $N = 2s + 1$ and values of $\ell$. We observe that the discrepancy between $A_\ell$ and $B_\ell$ is large for most $\ell$, except for small values ($\ell = 0, 1$). At first glance, this seems to contradict Fig.~\ref{fig:correlator}, where the position-space correlators rapidly approach their theoretical expectations. The resolution lies in the fact that the spherical harmonic expansion in Eq.~\eqref{eq:haromincexpansion} itself converges slowly in $\ell$. This slow convergence is especially evident in correlators between antipodal points, where the expansion becomes an alternating series $\sum_\ell (-1)^\ell$. To address this, it is useful to introduce a regularization factor $K_\ell = \sqrt{2s + 1} \tj{s}{s}{\ell}{s}{-s}{0}$, so that the correlator is instead expanded as
\begin{equation}\label{eq:regularizedharmonicexpansion}
\langle O(\bm{x}_1)\, O(\bm{x}_2) \rangle = \lim_{s \to \infty} \sum_{\ell = 0}^{2s} \sum_{m = -\ell}^{\ell} K_\ell A_\ell\, \bar{Y}_{\ell, m}(\bm{x}_1)\, Y_{\ell, m}(\bm{x}_2).
\end{equation}
This regularized series converges significantly faster with $s$ than the original expansion and, importantly, also properly regularizes divergent cases in the spherical harmonic expansion, such as the correlator $\langle \pi(\bm{x}_1)\, \pi(\bm{x}_2) \rangle$ discussed in Sec.~\ref{sec:freeQFT}.

Interestingly, the fuzzy sphere correlator appears to approximate the regularized version of the correlator. Fig.~\ref{fig:coefficients}(b) compares $B_\ell$ with the regularized coefficients $K_\ell A_\ell$, showing excellent agreement up to a cutoff scale $\ell \lesssim \sqrt{N}$. In the fuzzy sphere model, the emergence of the regularization factor $K_\ell$ can be traced to the integral
\begin{equation}
\int_{S^2} d^2\bm{x}\, \bar{Y}^s_{s, m_1}(\bm{x})\, Y^s_{s, m_2}(\bm{x})\, Y_{\ell, m}(\bm{x}), \nonumber
\end{equation}
which naturally introduces the $3j$ symbol structure $ \tj{s}{s}{\ell}{s}{-s}{0}$ appearing in $K_\ell$.

\subsection{Effective harmonic oscillators of fuzzy real scalar}

Since the real scalar CFT satisfies the harmonic oscillator algebra, it is interesting to investigate it in the context of our fuzzy sphere model. In the free scalar theory, the harmonic oscillator is
\begin{equation}
a_{\ell,m} = \frac{1}{2R} \left( \sqrt{\frac{2\ell+1}{R}} \phi_{\ell,m} + i \sqrt{\frac{4R}{2\ell+1}} \pi_{\ell,m} \right),
\end{equation} 
where $\phi_{\ell,m}$ and $\pi_{\ell,m}$ are the spherical modes of $\phi(\bm x)$ and $\pi(\bm x)$,
\begin{align}
\phi(\bm x) & = \frac{1}{R^2} \sum_{\ell=0}^\infty \sum_{m=-\ell}^\ell \phi_{\ell,m} Y_{\ell,m}(\bm x), \\ 
\pi(\bm x) & = \frac{1}{R^2} \sum_{\ell=0}^\infty \sum_{m=-\ell}^\ell \pi_{\ell,m} Y_{\ell,m}(\bm x).
\end{align}
Similarly, we introduce the spherical modes of fields on the fuzzy sphere,
\begin{equation}\label{eq:fuzzydensity}
n^A(\bm x) =  \frac{1}{2s+1} \sum_{\ell=0}^{2s}\sum_{m=-\ell}^{\ell} n^A_{\ell,m} Y_{\ell,m}(\bm x).
\end{equation}
Given the correspondence $n^x(\bm{x}) \sim \phi(\bm{x})$ and $n^y(\bm{x}) \sim \pi(\bm{x})$, it is natural to ask if the superposition of  $n^{x,y}_{\ell,m}$ gives the harmonic oscillator,
\begin{equation} \label{eq:effa}
\tilde{a}_{\ell,m} = f_{\ell} \, n^x_{\ell,m} - i g_{\ell} \, n^y_{\ell,m}.
\end{equation}
Here, we include a tilde to $a_{\ell,m}$ to avoid the potential confusion with the notation used in the free scalar theory.

We can first determine $ f_\ell$ and $ g_\ell$ numerically from the ground state $|\bm 0\rangle$. For the true harmonic oscillators we have,
\begin{equation}
\langle \bm 0 | \tilde a_{\ell, m} \tilde a_{\ell,m}^\dag |\bm 0 \rangle = 1, \quad \langle \bm 0 | \tilde a_{\ell, m}^\dag \tilde a_{\ell,m} |\bm 0 \rangle =0.
\end{equation}
In practice, the second condition can almost never be satisfied at a finite $s$. So we determine $ f_\ell$ and $ g_\ell$ by minimizing $\langle\bm 0 | \tilde a_{\ell, m}^\dag \tilde a_{\ell,m} |\bm 0 \rangle$ under the condition that $\tilde a^\dag_{\ell,m}|\bm 0\rangle$ is normalized. This can be solved by a generalized eigenvalue problem as detailed in the Appendix~\ref{sec:semidefinite}. 

Numerically, we find that the coefficients $(f_{\ell}, g_{\ell})$ computed from the ground state closely match the analytical expressions (as shown in Fig.~\ref{fig:coefficients}),
\begin{equation}
f_{\ell}  = \frac{1}{2R'K_{\ell}} \sqrt{\frac{(2\ell+1)K_{\ell}}{R'}}, \quad
g_{\ell}  = \frac{1}{2R'K_{\ell}} \sqrt{\frac{R'}{(2\ell+1)K_{\ell}}},
\end{equation}
where $K_{\ell} = \sqrt{2s+1}\tj{s}{s}{\ell}{s}{-s}{0}$ is the regularization factor, and $R' = 0.95\sqrt{2s+1} \approx \sqrt{|\langle n^z_{00}\rangle|}$. Discrepancies begin to appear at $\ell \sim \sqrt{N}$, corresponding to the cutoff scale of the fuzzy sphere model. Interestingly, the $(f_\ell, g_\ell)$ obtained on the fuzzy sphere differ from those in standard QFT only by the regularization factor $K_\ell$. As we have discussed, this factor facilitates the convergence of correlators, and it is intriguing that the fuzzy sphere model naturally incorporates it.

\begin{figure}
    \centering
    \includegraphics[width=0.495\linewidth]{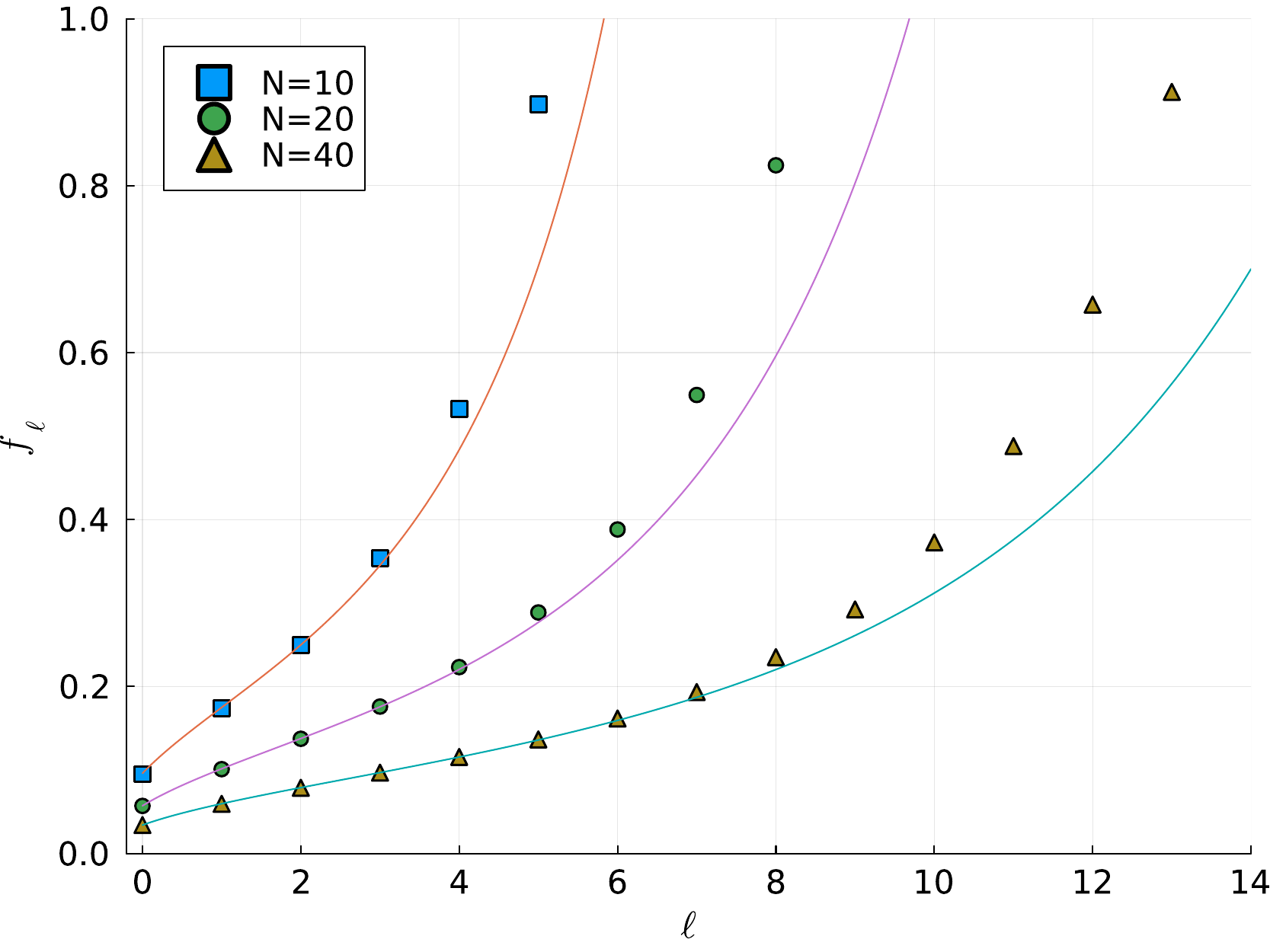}
        \includegraphics[width=0.495\linewidth]{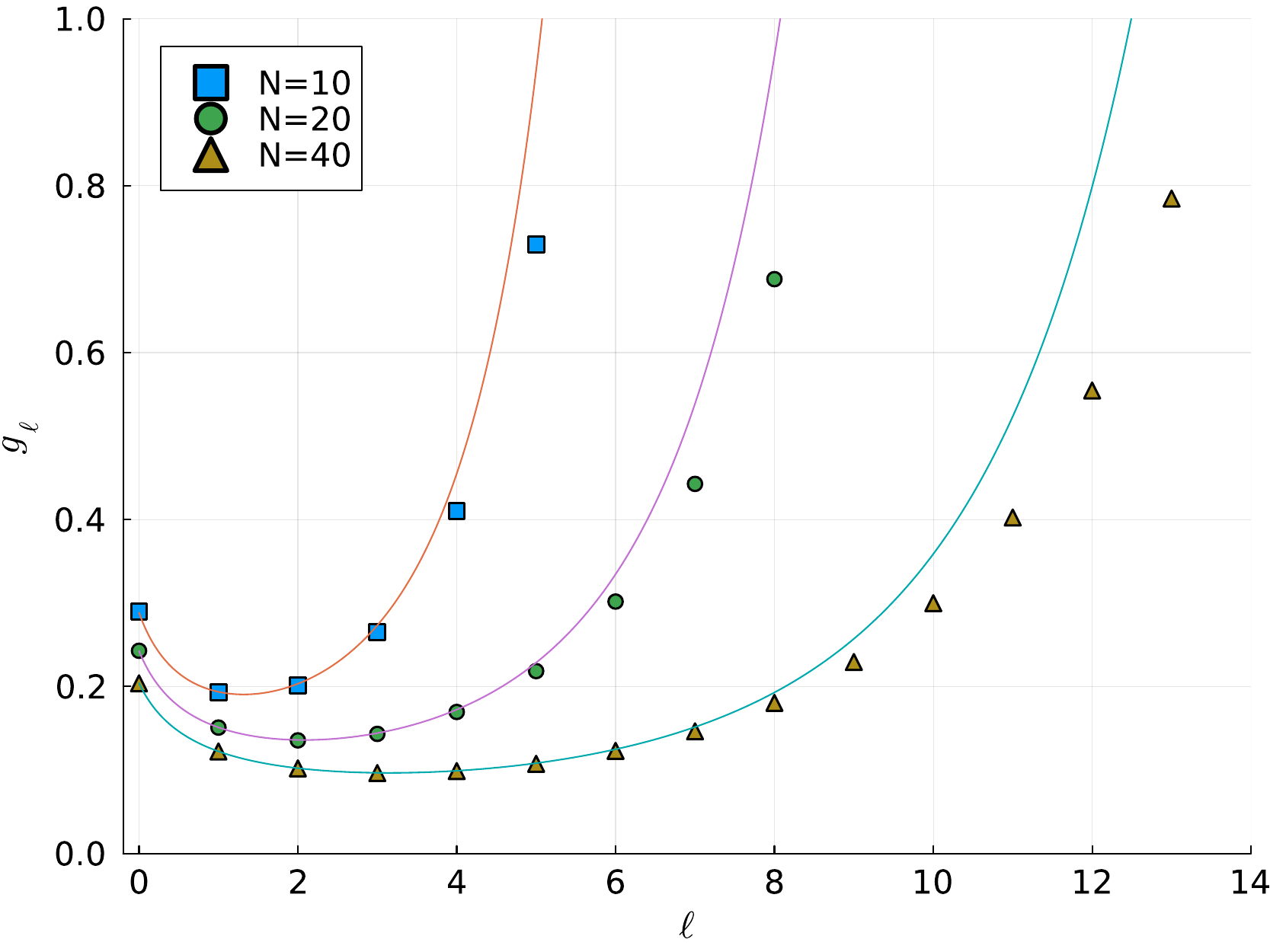}
    \caption{The numerical coefficients of the effective harmonic oscillator $\tilde{a}_{\ell,m} = f_{\ell} \, n^x_{\ell,m} - i g_{\ell} \, n^y_{\ell,m}$ closely follow the analytical expressions: $f_{\ell}  = \frac{1}{2R'K_{\ell}} \sqrt{\frac{(2\ell+1)K_{\ell}}{R'}}$,
 $g_{\ell}  = \frac{1}{2R'K_{\ell}} \sqrt{\frac{R'}{(2\ell+1)K_{\ell}}}$.}
    \label{fig:harmonic_coeff}
\end{figure}

\begin{figure}
    \centering
    \includegraphics[width=0.495\linewidth]{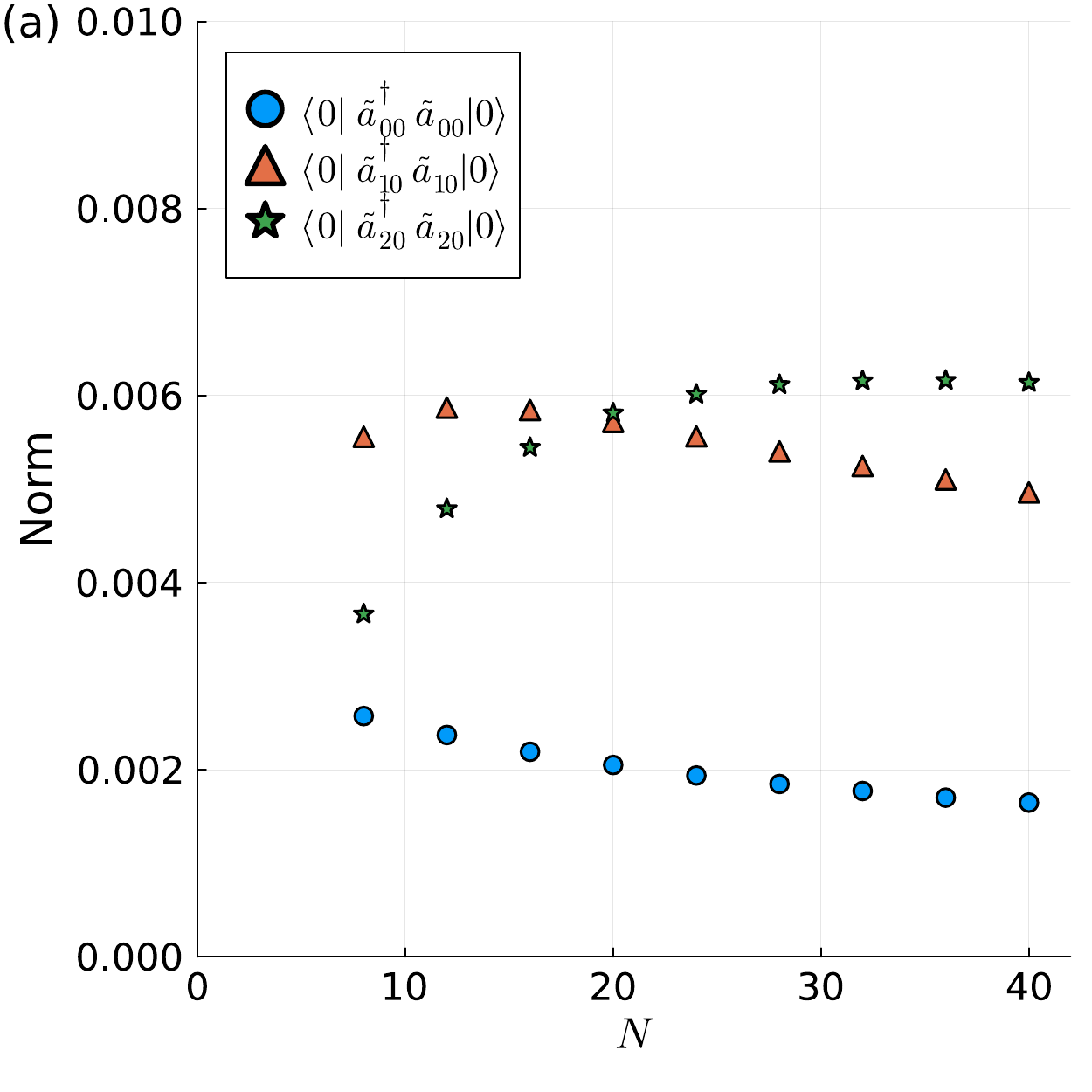}
        \includegraphics[width=0.495\linewidth]{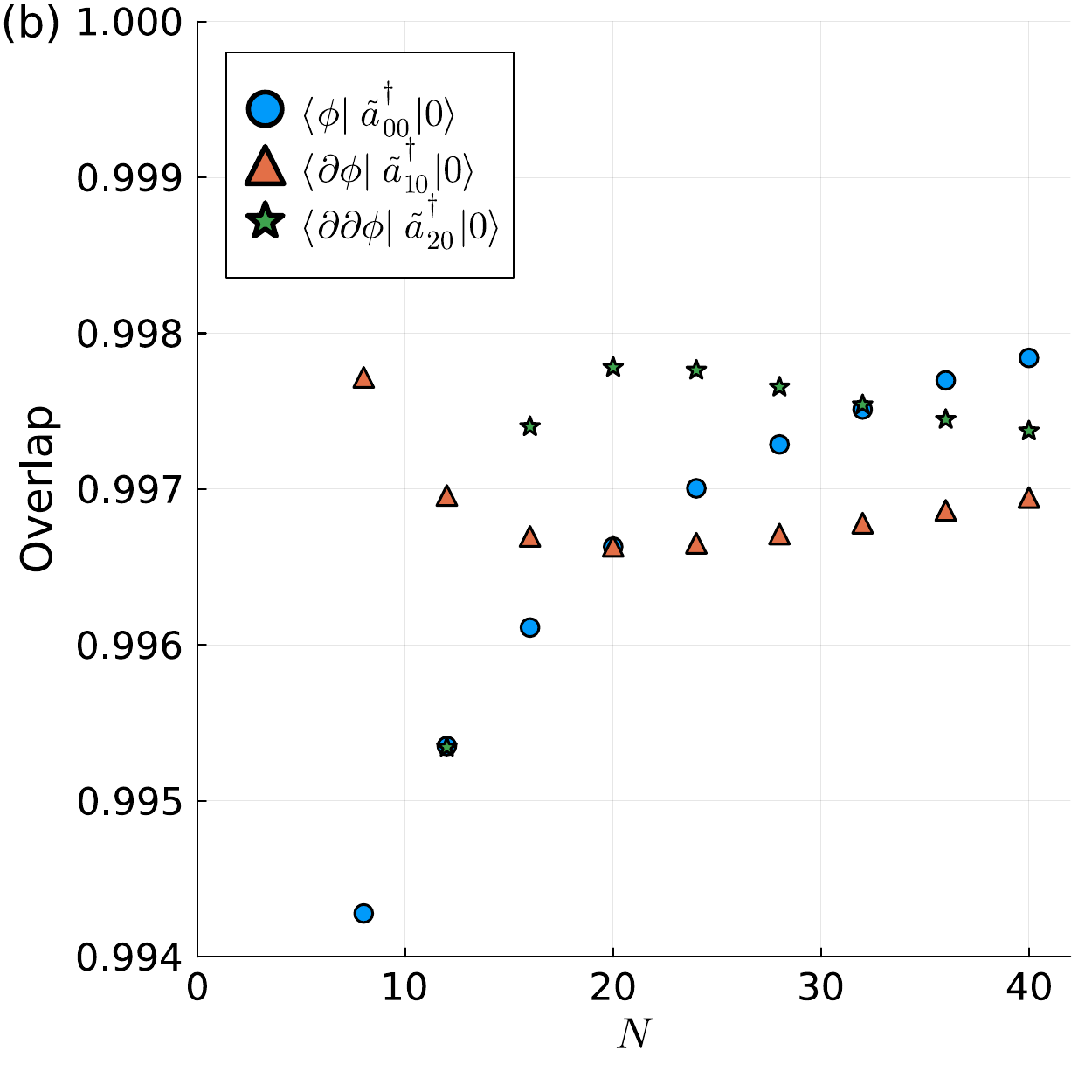}
    \caption{The quality of effective harmonic oscillators on fuzzy sphere: the norm $\langle 0 | \tilde a_{\ell,m}^\dag a_{\ell,m} |0\rangle $ is close to $0$ and $a_{\ell,m}^\dag  |0\rangle $ creates the CFT state $|\phi\rangle$ and its descendants. }
    \label{fig:Harmonic}
\end{figure}

\begin{figure}
    \centering
    \includegraphics[width=0.495\linewidth]{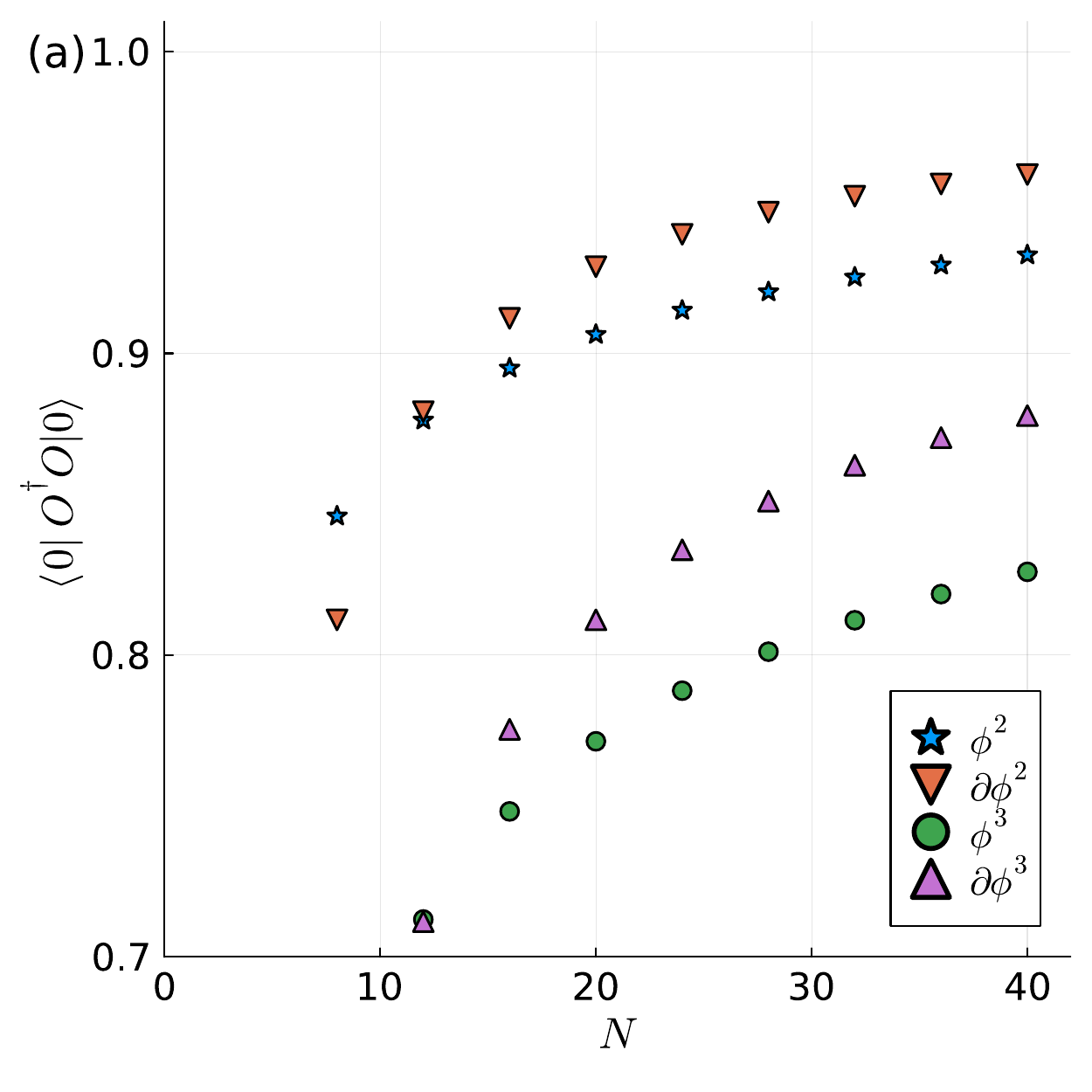}
        \includegraphics[width=0.495\linewidth]{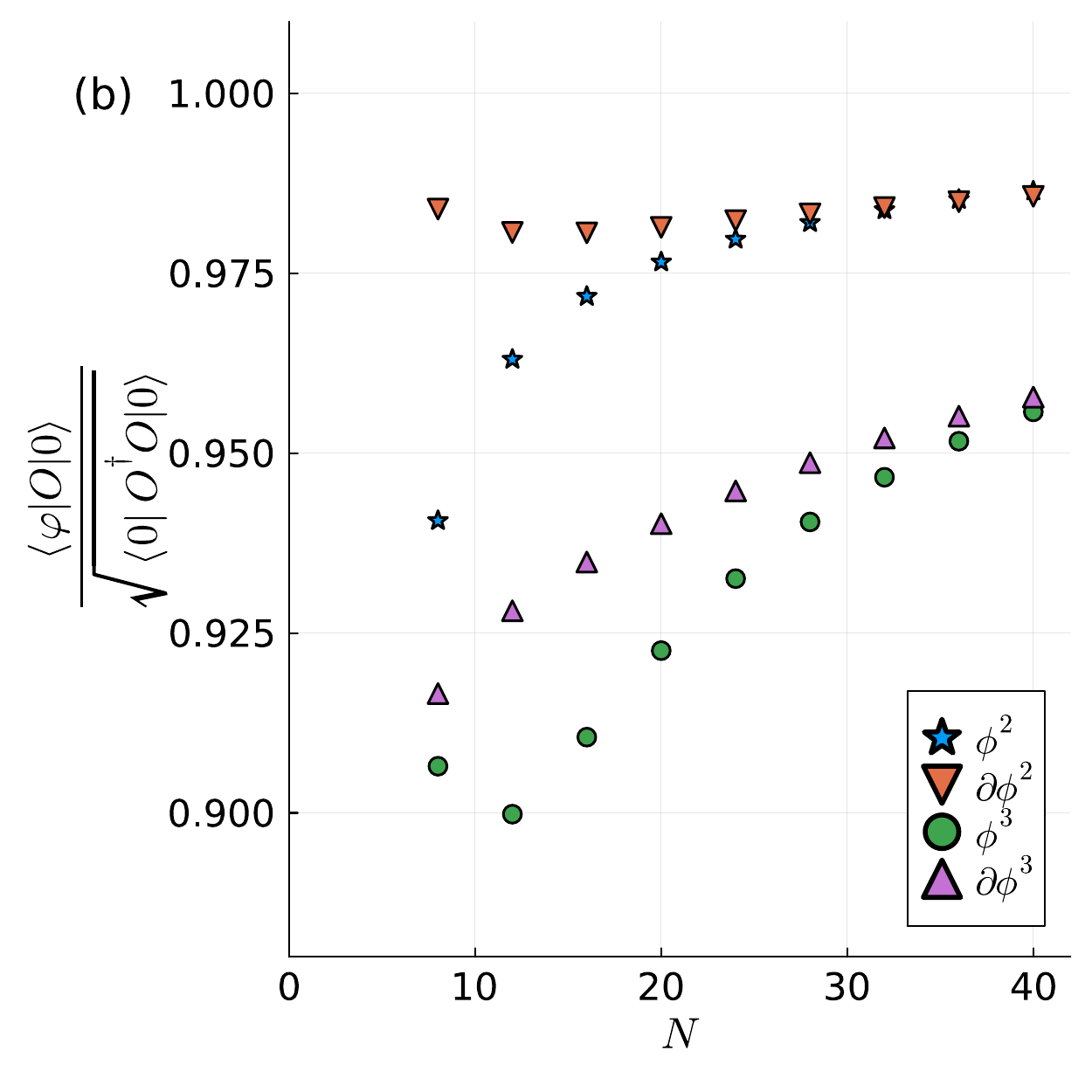}
    \caption{Creating higher CFT states using effective harmonic oscillators on fuzzy sphere.}
    \label{fig:Harmonic2}
\end{figure}

Fig.~\ref{fig:Harmonic} illustrates the quality of our constructed harmonic oscillators for small $\ell$. First, we observe that $\tilde a_{\ell,m}$ approximately annihilates the ground state, i.e., $\langle \bm 0|\tilde a_{\ell,m}^\dag \tilde a_{\ell,m}|\bm 0\rangle \approx 0$, as shown in Fig.~\ref{fig:Harmonic}(a). On the other hand, applying $\tilde a_{\ell,m}^\dag$ to the ground state should generate CFT states. For example, $\tilde a_{0,0}^\dag |\bm 0\rangle$ produces the lowest primary state $|\phi\rangle$, while $\tilde a_{\ell,m}^\dag |\bm 0\rangle$ yields the spin-$\ell$ descendant of $|\phi\rangle$. Fig.~\ref{fig:Harmonic}(b) displays the overlaps between the CFT states (i.e., the eigenstates of our fuzzy sphere Hamiltonian) and $\tilde a_{\ell,m}^\dag |\bm 0\rangle$, which are remarkably close to $1$. Furthermore, we examine higher excited states,
\begin{equation}\label{eq:harmonichigher}
|\phi^2\rangle = \frac{(a^\dag_{0,0})^2}{2}|\bm 0\rangle, \quad 
|\partial \phi^2\rangle = a^\dag_{1,m} a^\dag_{0,0} |\bm 0\rangle, \quad 
|\phi^3\rangle = \frac{(a^\dag_{0,0})^3}{3!}|\bm 0\rangle, \quad 
|\partial \phi^3\rangle = \frac{a^\dag_{1,m} (a^\dag_{0,0})^2}{2}|\bm 0\rangle.
\end{equation}
We compare the CFT states obtained on the fuzzy sphere, $|\varphi\rangle = |\phi^2\rangle, |\partial \phi^2\rangle, |\phi^3\rangle, |\partial \phi^3\rangle$, with the corresponding states constructed via the effective harmonic oscillators, $O|\bm 0\rangle$, where $O = (\tilde a^\dag_{0,0})^2/2$, $\tilde a^\dag_{1,m} \tilde a^\dag_{0,0}$, $(\tilde a^\dag_{0,0})^3/3!$, and $\tilde a^\dag_{1,m} (\tilde a^\dag_{0,0})^2/2$, respectively. We note that $O|\bm 0\rangle$ is not automatically normalized, so its norm serves as a nontrivial consistency check for our constructed fuzzy sphere harmonic oscillators. As shown in Fig.~\ref{fig:Harmonic2}(a), the norm $\langle \bm 0|O^\dag O|\bm 0\rangle$ is generally less than $1$ due to finite-size effects, but it approaches $1$ as the system size $N$ increases. Moreover, the (normalized) overlaps between $|\varphi\rangle$ and $O|\bm 0\rangle$ are close to $1$ and improve with increasing system size. These observations defy the generic expectation of orthogonality catastrophe in quantum many-body systems, where the overlap between different states typically decays exponentially to zero. Therefore, our results strongly support the emergence of a free scalar CFT and affirm the validity of the constructed effective harmonic oscillators on the fuzzy sphere.

\subsection{Shift symmetry and non-compact scalar}\label{sec:shiftsymmetry}
We now discuss how the $U(1)$ symmetry of our fuzzy sphere model becomes the shift ($\mathbb{R}$) symmetry of the real scalar field. The underlying mechanism is the familiar one associated with the superfluid phase, namely $U(1)$ spontaneous symmetry breaking, which we briefly review first. In the case of a superfluid, one starts with a complex scalar field $\Phi(\bm x)$ and introduces a phase field $\theta(\bm x)$ via $\Phi(\bm x) = \rho(\bm x) e^{i\theta(\bm x)}$. Under a $U(1)$ phase rotation, $\Phi(\bm x) \rightarrow e^{ic} \Phi(\bm x)$, the phase field transforms as a shift symmetry: $\theta(\bm x) \rightarrow \theta(\bm x) + c$. The amplitude field $\rho(\bm x)$ is massive, so the effective low energy theory can be described by fluctuations of the phase field $\theta(\bm x)$,
\begin{equation}
\mathcal{L} = \frac{f}{2} (\partial \theta(\bm x))^2 + \textrm{higher-order derivative terms...}
\end{equation}
This Lagrangian resembles that of a standard real scalar field theory, with one important distinction: $\theta(\bm x)$ is a compact boson, i.e., $\theta(\bm x) = \theta(\bm x) + 2\pi$. The compactness of $\theta(\bm x)$ arises from the presence of vortex configurations in the path integral. In the $2+1$-dimensional case considered here, a vortex forms a string-like object, and we have the quantization condition $\int_{S^1} d\bm x\, \partial \theta(\bm x) = 2\pi$ for a loop winds around the vortex. Consequently, $\theta(\bm x)$ must be multivalued as $\theta(\bm x) = \theta(\bm x) + 2\pi$; otherwise, it would exhibit discontinuities in spacetime.

Returning to our fuzzy sphere model, the real scalar field $\phi(\bm x)$ is similarly related to the fuzzy sphere field via
\begin{equation}
n^z(\bm x) + i n^x(\bm x) \sim e^{i\phi(\bm x)} (1 + \textrm{higher-order derivative terms...}),
\end{equation}
such that the $U(1)$ rotation generated by $n^y_{00}$ corresponds to a shift symmetry of $\phi(\bm x)$. However, in our case, the scalar field $\phi(\bm x)$ is non-compact due to vortex confinement induced by the $U(1)$-breaking term $\frac{h}{R^2} n^z(\bm x)$ in the Hamiltonian. When $h = 0$, the model will be in the superfluid phase\footnote{This also explains the choice of sign for the $n^y(\bm x) \nabla^2 n^y(\bm x)$ term in our Hamiltonian Eq.~\eqref{eq:fuzzy_gaussiantext}. An opposite sign will lead to the $\mathbb Z_2$ ferromagnet of $n^y$.}. The $U(1)$-breaking term energetically favors aligning the isospin along the $n^z(\bm x)$ direction, thereby pinning the vacuum and inducing a mass for the scalar field described by the sine-Gordon term $m^2 \cos(\phi(\bm x))$, where $m = \sqrt{h}/R$. Crucially, $\phi(\bm x)$ becomes non-compact due to the linear confinement of vortices caused by the sine-Gordon term, even though its strength scales to zero as $1/R^2$. To understand this, consider a vortex–antivortex pair separated by a distance $L$. The total energy of such a configuration is of order $m L \beta = \sqrt{h} L \beta / R$, where $\beta$ is the inverse temperature (i.e., energy scale). In the low-energy regime of interest, we have $\beta \sim R$, implying that the energy of the vortex pair scales as $\sqrt{h} L$. Hence, vortices experience linear confinement, rendering $\phi(\bm x)$ effectively non-compact.

Therefore, the low energy theory of model is a non-compact scalar described by the effective theory,
\begin{equation}
\mathcal L = \frac{f}{2}(\partial \phi(\bm x))^2 + f_1(\partial \phi(\bm x))^4 + \cdots + \frac{h'}{2R^2}(\phi^2(\bm x) -  \phi^4(\bm x)/12 + \cdots),
\end{equation}
where all the shift symmetry breaking terms (e.g. $\phi^2(\bm x)$, $\phi^4(\bm x)$) appear with $1/R^2$. Therefore, for the free real scalar fixed point, our model only has a marginal perturbation $\phi^2(\bm x)/R^2$ that we need to fine tune. In practice, we have an extra fine tuning to kill the next irrelevant operator to minimize the finite size effect.
 
\section{Fuzzy sphere algebra, wave-function ansatz and Ising CFT}
\label{sec:fuzzyalgebra}
\subsection{Fuzzy sphere algebra and harmonic oscillator}
As we have shown, the density operators $n^{x,y}_{\ell, m}$ can give accurate approximations of the harmonic oscillators in the free scalar CFT. This motivates a direct study of the algebra satisfied by these density operators introduced in Eq.~\eqref{eq:fuzzydensity}. These density operators are explicitly written as, 
\begin{align}
	n_{\ell,m}^A 
	&= (2s+1)\sqrt{2 \ell+1} \tj{s}{s}{\ell}{s}{-s}{0} \sum_{m_1=-s}^s (-1)^{s+m_1}   \tj{s}{s}{\ell}{m_1}{m-m_1}{-m} 
	c^\dag_{m_1,\alpha} c_{m_1-m,\beta} A_{\alpha\beta}.
 \end{align}
 Here $A$ is a $2\times 2$ matrix, and for $n^{x,y,z,0}$ it corresponds to the Pauli matrix $\sigma^{x,y,z}$ and identity matrix. Also apparently we have  $n^{\alpha A}_{\ell,m} = \alpha n^A_{\ell,m}$ for any $\mathbb C$-number $\alpha$. The density operators have a natural truncation at $\ell=2s$, and they satisfy the following algebra: 
\begin{align}\label{eq:fuzzyalgebra}
 [n_{\ell_a,m_a}^A, n_{\ell_b,m_b}^B]  & = \sum_{ \substack{\ell=0 \\ \ell+\ell_a+\ell_b\in \textrm{even}} }^{2s} h_\ell(s, \ell_a,\ell_b) (-1)^{m_a+m_b} \tj{\ell_a}{\ell_b}{\ell}{m_a}{m_b}{-m_a-m_b}   n_{\ell,m_a+m_b}^{[A,B]} \nonumber \\
 & -  \sum_{ \substack{\ell=0 \\ \ell+\ell_a+\ell_b\in \textrm{odd}} }^{2s} h_\ell(s, \ell_a,\ell_b) (-1)^{m_a+m_b} \tj{\ell_a}{\ell_b}{\ell}{m_a}{m_b}{-m_a-m_b}   n_{\ell,m_a+m_b}^{\{A,B\}},
\end{align}
where 
\begin{equation}
h_\ell(s, \ell_a,\ell_b) =  (2s+1)\sqrt{(2\ell_a+1)(2\ell_b+1)(2\ell+1)} \sj{\ell_a}{\ell_b}{\ell}{s}{s}{s} \frac{ \tj{s}{s}{\ell_a}{s}{-s}{0}  \tj{s}{s}{\ell_b}{s}{-s}{0} }{\tj{s}{s}{\ell}{s}{-s}{0}},
\end{equation}
 $(\cdots)$ and $\{\cdots \}$ are $3j$ and $6j$ symbols. 

This algebra is closely related to the GMP (Girvin–MacDonald–Platzman) algebra~\cite{GMP}, which has been extensively studied in the condensed matter literature. The GMP algebra is primarily formulated in flat space and has played a central role in understanding the magneto-roton excitations of fractional quantum Hall effect in the lowest Landau level. It is also noteworthy that the GMP algebra in flat space is equivalent to the star-product structure used in non-commutative geometry~\footnote{In flat space, GMP algebra considers density operators $n_{\bm k}$ which follows the algebra  $ [n_{\vec q}, n_{\vec k}] = 2 i \sin\left[ \frac{l^2_B}{2} |\vec q \times \vec k|  \right] n_{\vec q+ \vec k}$, where $l_B$ is the magnetic length. This is exactly the Moyal bracket, i.e., antisymmetrized star product used in non-commutative field theory. }. In contrast to the conventional GMP algebra, our setup considers a spherical geometry and, more importantly, incorporates an additional $SU(2)$ flavor (isospin) degree of freedom which enriches the algebra.

To gain more intuition about this algebra, let us first consider the case without the $SU(2)$ flavor degree of freedom, i.e., the spherical GMP algebra. In this case, the algebra takes the form
\begin{align}
[n_{\ell_a,m_a}, n_{\ell_b,m_b}] = - \sum_{ \substack{\ell=0 \\ \ell+\ell_a+\ell_b \in \textrm{odd}} }^{2s} 2h_\ell(s, \ell_a, \ell_b) (-1)^{m_a + m_b} \tj{\ell_a}{\ell_b}{\ell}{m_a}{m_b}{-m_a - m_b} n_{\ell, m_a + m_b}.
\end{align}
The simplest representation of the spherical GMP algebra is given by the fuzzy spherical harmonics, which are $(2s+1) \times (2s+1)$ matrices. These harmonics form the foundational basis for non-commutative field theory on the fuzzy sphere~\cite{fuzzypathintegral,Madore2001Scaling}. Physically, this representation corresponds to a single fermion occupying the lowest Landau level. Similarly, one can consider the case with $k \leq 2s+1$ fermions in the lowest Landau level, leading to a representation of dimension $\binom{2s+1}{k}$. 
It remains unclear whether these representations are irreducible, and it is likely that there exist other irreducible representations not directly related to Landau level physics. For example, when $s = 1/2$, the algebra reduces to the $SU(2)$ algebra, which admits an infinite number of representations, only one of which—the $j = 1/2$ representation—is connected to the Landau level.
It would be valuable to identify the full set of Casimir operators of this algebra. An obvious Casimir is $n_{0,0}$, which, in the context of the Landau level problem, corresponds to the fermion number. A nontrivial Casimir we have found is
\begin{equation}
Q = \sum_{\ell=0}^{2s} \sum_{m=-\ell}^\ell \frac{(-1)^m n_{\ell,m} n_{\ell,-m}}{\tj{s}{s}{\ell}{s}{-s}{0}^2},
\end{equation}
which, in the case of $s=1/2$, reduces to the $SU(2)$ Casimir operator. For the representations associated with the lowest Landau level, this Casimir takes the value
\begin{equation}
Q = (-N^2 - N + 1)\, n_{0,0}^2 + N^2(N+1)\, n_{0,0},
\end{equation}
where $N = 2s + 1$. This result can be derived by explicitly expressing $n_{\ell,m}$ in terms of fermionic operators on the lowest Landau level.

Let us now return to the fuzzy sphere algebra in Eq.~\eqref{eq:fuzzyalgebra}. An important observation is that, in a certain limit, the fuzzy sphere algebra reduces to that of harmonic oscillators~\footnote{We thank Davide Gaiotto for discussion.}. To make this connection more transparent, we  define the $n^{\pm}_{\ell,m} = \frac{1}{2} \left(n^x_{\ell,m} \pm i n^y_{\ell,m}\right)$, whose commutation relations are given by
\begin{align}
[n^+_{\ell_a,m_a}, n^+_{\ell_b,m_b}] &= [n^-_{\ell_a,m_a}, n^-_{\ell_b,m_b}] = 0, \\
[n^+_{\ell_a,m_a}, n^-_{\ell_b,m_b}] &= \sum_{\substack{\ell=0 \\ \ell+\ell_a+\ell_b \in \textrm{even}}}^{2s} h_\ell(s, \ell_a, \ell_b) (-1)^{m_a + m_b} \tj{\ell_a}{\ell_b}{\ell}{m_a}{m_b}{-m_a - m_b} n^z_{\ell, m_a + m_b} \nonumber \\
&\quad - \sum_{\substack{\ell=0 \\ \ell+\ell_a+\ell_b \in \textrm{odd}}}^{2s} h_\ell(s, \ell_a, \ell_b) (-1)^{m_a + m_b} \tj{\ell_a}{\ell_b}{\ell}{m_a}{m_b}{-m_a - m_b} n^0_{\ell, m_a + m_b}.
\end{align}
We consider a reference state,
$|\varphi_0\rangle = \prod_{m=-s}^s c^\dagger_{m,\downarrow} |\bm 0\rangle$, which consists of exactly one spin-down fermion occupying each Landau orbital. It is straightforward to verify that $n^-_{\ell,m}$ annihilates $|\varphi_0\rangle$, while $n^+_{\ell,m}$ creates non-trivial states when acting on $|\varphi_0\rangle$. We now focus on a subspace spanned by the states
\begin{equation} \label{eq:subspace}
\prod_{i=1}^k n^+_{\ell_i,m_i} |\varphi_0\rangle, \qquad \ell_i \ll s, \quad k \ll s.
\end{equation}
Within this subspace, the following approximations hold:
\[
n^z_{\ell=0,m=0} \sim 2s + 1 + \mathcal{O}(s^0), \qquad 
n^z_{\ell \neq 0,m} \sim \mathcal{O}(s^0), \qquad 
n^0_{\ell \neq 0,m} \sim \mathcal{O}(s^0).
\]
Therefore, in the infinite $s$ limit, the rescaled operators $n^+_{\ell,m}/\sqrt{2s+1}$ and $n^-_{\ell,m}/\sqrt{2s+1}$ approximately obey the canonical commutation relations of harmonic oscillator creation and annihilation operators:
\begin{equation}
\left[\frac{n^+_{\ell_a,m_a}}{\sqrt{2s+1}}, \frac{n^-_{\ell_b,m_b}}{\sqrt{2s+1}}\right] \approx \delta_{\ell_a,\ell_b} \delta_{m_a,m_b} \frac{n^z_{\ell=0,m=0}}{2s+1} \approx \delta_{\ell_a,\ell_b} \delta_{m_a,m_b}.
\end{equation}
We remark there is a subtlety that, although each $n^z_{\ell \neq 0,m}/(2s+1)$ and $n^0_{\ell \neq 0,m}/(2s+1)$ vanishes individually in the infinite $s$ limit, there are $\sim s$ such terms, and their cumulative contribution may not be negligible. We will leave a more rigorous analysis for future work.

The harmonic oscillator approximation is valid only within the restricted subspace defined in Eq.~\eqref{eq:subspace}, 
\footnote{The free scalar CFT and the Ising CFT realized on the fuzzy sphere do not live in this subspace. One way to see this is that the expectation value of $n^z_{0,0}/(2s+1)$ is less than $1$ for the ground states of both the free scalar and Ising CFTs.}
and only for density operators $n^\pm_{\ell,m}$ with $\ell \ll s$. Therefore, in the fuzzy sphere model, one should not naively treat the density operators as harmonic oscillators. One reason is that the Hamiltonian in Eq.~\eqref{eq:fuzzy_Ising} consists only of bilinears of the density operators, yet in the continuum limit $s \rightarrow \infty$, it realizes the interacting 3D Ising CFT. Even in the case of the free scalar CFT, if one naively interprets the density operators as harmonic oscillators, the Hamiltonian in Eq.~\eqref{eq:fuzzy_gaussiantext} would take the form
\begin{equation}
H = \int_{S^2} d^2 \bm x \left( U\, \pi(\bm x) \nabla^2 \pi(\bm x) + \frac{h}{R^2} \phi^2(\bm x) \right),
\end{equation}
which differs significantly from the standard free scalar QFT given in Eq.~\eqref{eq:freescalar}.

There are two key fundamental differences between the fuzzy density operator $n^A_{\ell,m}$ and the harmonic oscillator $a_{\ell,m}$. It is best to understand these in terms of their real space counterparts, $n^{x,y,z,0}(\bm x)$ and $\phi(\bm x)$, $\pi(\bm x)$. 

First, in standard QFT, one can construct composite fields from the fundamental field $\phi(\bm x)$, such as $\phi^2(\bm x)$, $\phi^3(\bm x)$, $\phi(\bm x) \partial_\mu \phi(\bm x)$, and so on. Similarly, in the fuzzy sphere model, composite fields can be constructed schematically from $n^A(\bm x)$. However, due to the finite-dimensional Hilbert space and algebraic constraints, some of these operators become equivalent. For instance, we find the relation
\begin{equation}
\left(n^x(\bm x)\right)^2 = \left(n^y(\bm x)\right)^2 = \left(n^z(\bm x)\right)^2 = 2n^0(\bm x) - \left(n^0(\bm x)\right)^2.
\end{equation}
This identity can be understood by evaluating the fields at the north pole, where
\begin{equation}\label{eq:northpole}
n^A(\bm x = \text{North pole}) = \left(c^\dagger_{s,\uparrow}, c^\dagger_{s,\downarrow}\right) A 
\begin{pmatrix}
c_{s,\uparrow} \\
c_{s,\downarrow}
\end{pmatrix}.
\end{equation}
This follows from the structure of the monopole harmonics, specifically the fact that only $Y_{s,s}^{(s)}(\bm x)$ is nonvanishing at the north pole.

Secondly, the fuzzy sphere model is free from the UV divergences that are commonly encountered in standard QFT. This feature is consistent with one of the original motivations for non-commutative geometry in physics, as proposed by Heisenberg. In conventional QFT, UV divergences can often be traced back to Eqs.~\eqref{eq:phifield}–\eqref{eq:pifield}, where the quantum fields involve infinite sums over modes. Naively, one might expect that the fuzzy sphere model would suffer from a similar divergence due to the infinite sum (as $s \rightarrow \infty$) in Eq.~\eqref{eq:fuzzydensity}, especially considering that the fuzzy sphere algebra reduces to that of harmonic oscillators in certain limits. However, in the fermionic representation—particularly when evaluated at the north pole—the density field reduces to a finite, non-divergent expression as shown in Eq.~\eqref{eq:northpole}. Moreover, one can explicitly show that the correlators remain non-divergent. As an example, consider the two-point correlator,
\begin{align}
& \langle n^A(\bm x=\text{North pole})n^A(\bm x) \rangle = \sum_{m=-s}^s \langle A_{\alpha,\beta} A_{\eta,\gamma} c_{s,\alpha}^\dag c_{s,\beta} c_{m,\eta}^\dag c_{m,\gamma}\rangle \bar{Y}_{s,m}^{(s)}(\bm x) Y_{s,m}^{(s)}(\bm x) \nonumber \\ &= \sum_{m=-s}^s \langle A_{\alpha,\beta} A_{\eta,\gamma} c_{s,\alpha}^\dag c_{s,\beta} c_{m,\eta}^\dag c_{m,\gamma}\rangle \binom{2s}{s+m} \cos^{2s+2m}\left(\frac{\theta}{2}\right)\sin^{2s-2m}\left(\frac{\theta}{2}\right) \nonumber    \\
& \le \sum_{m=-s}^s \binom{2s}{s+m} \cos^{2s+2m}\left(\frac{\theta}{2}\right)\sin^{2s-2m}\left(\frac{\theta}{2}\right) = 1.
\end{align}
Similarly, one can prove other composite fields are also UV finite.

\subsection{Wavefunction ansatz}

In this section, we propose a wavefunction ansatz for the ground state of the free scalar CFT on the fuzzy sphere. As we show numerically, the ground state is annihilated by the effective harmonic oscillator operator defined in Eq.~\eqref{eq:effa}:
\begin{align}
\tilde a_{\ell,m} &= \frac{1}{2}\left( \sqrt{\frac{R'}{(2\ell+1)K_\ell}} + \sqrt{\frac{(2\ell+1)K_\ell}{R'}} \right) \frac{n^-_{\ell,m}}{R' K_\ell} 
- \frac{1}{2}\left( \sqrt{\frac{R'}{(2\ell+1)K_\ell}} - \sqrt{\frac{(2\ell+1)K_\ell}{R'}} \right) \frac{n^+_{\ell,m}}{R' K_\ell} \nonumber \\
&= u_\ell \frac{n^-_{\ell,m}}{R' K_\ell} - v_\ell \frac{n^+_{\ell,m}}{R' K_\ell},
\end{align}
where $R' = 0.95\sqrt{2s+1} \approx \sqrt{|\langle n^z_{00} \rangle|}$. The operators $n^\pm_{\ell,m}/(R' K_\ell)$ are approximately creation and annihilation operators of harmonic oscillators in an appropriate limit, and the corresponding vacuum is given by $\prod_{m=-s}^s c^\dagger_{m,\downarrow} |\bm 0\rangle$. Thus, $\tilde a_{\ell,m}$ can be viewed as a Bogoliubov quasiparticle operator, especially $u_\ell$ and $v_\ell$ satisfy the relation $u_\ell^2 - v_\ell^2 = 1$.
Motivated by this, we propose the following ansatz for the ground state wavefunction of the free real scalar CFT on the fuzzy sphere:
\begin{align}
|\psi_{\textrm{free ansatz}}\rangle &= \exp\left( \sum_{\ell=0}^{2s} \frac{1}{4} \log\left( \frac{u_\ell - v_\ell}{u_\ell + v_\ell} \right) \sum_{m=-\ell}^\ell (-1)^m \frac{n^-_{\ell,m} n^-_{\ell,-m} - n^+_{\ell,m} n^+_{\ell,-m}}{2s+1} \right) |\varphi_0\rangle \nonumber \\
&= \exp\left( \sum_{\ell=0}^{2s} \frac{1}{4} \log\big( (2\ell+1) K_\ell \big) \sum_{m=-\ell}^\ell (-1)^m \frac{n^-_{\ell,m} n^-_{\ell,-m} - n^+_{\ell,m} n^+_{\ell,-m}}{2s+1} \right) |\varphi_0\rangle. \label{eq:ansatz_free}
\end{align}
Naively, one may consider modifying $n^\pm_{\ell,m}$ by including an additional factor such as $1/K_\ell$ or $1/(0.95\, K_\ell)$, but numerical tests show that such modifications do not reproduce the correct ground state wavefunction. This may be due to that $n^\pm_{\ell,m}/(R' K_\ell)$ are not strictly harmonic oscillators. 

Inspired by the free scalar ansatz, we also propose a wavefunction ansatz for the Ising CFT, 
\begin{align}
|\psi_{\textrm{Ising ansatz}}\rangle 
&= \exp\left( \sum_{\ell=0}^{2s} \frac{1}{4} \log\left( \frac{2(1-\Delta)_{\ell+1}}{(\Delta)_\ell} K_\ell \right) \sum_{m=-\ell}^\ell (-1)^m \frac{n^-_{\ell,m} n^-_{\ell,-m} - n^+_{\ell,m} n^+_{\ell,-m}}{2s+1} \right) |\varphi_0\rangle, \label{eq:ansatz_Ising}
\end{align}
where $(\Delta)_\ell$ is the Pochhammer symbol and $\Delta\approx 0.518149$ is the scaling dimension of $\sigma$ primary operator of the $3D$ Ising CFT. When $\Delta=1/2$, this wavefunction reduces to the free scalar ansatz. 

Since the wavefunction ansatz involves an exponential of four-fermion operators, its computation is nontrivial. To evaluate it, we represent the wavefunction using matrix product states (MPS) and employ the time-dependent variational principle (TDVP) algorithm~\cite{TDVP2011,TDVP_White}. The first quantity used to assess the quality of the ansatz is its wavefunction overlap with the CFT ground state on the fuzzy sphere. As shown in Table~\ref{tab:overlap}, the overlap is remarkably close to $1$. Interestingly, the Ising CFT ansatz achieves an even higher overlap than the free scalar CFT ansatz, exceeding $0.99$ for systems with $N=28$ fermions. These overlaps are notably larger than the typical overlap between the Laughlin state~\cite{Laughlin} and the ground state of the partially filled lowest Landau level with Coulomb interaction~\cite{Laughlin_overlap}. Nevertheless, it is important to emphasize that, unlike the Laughlin state for the fractional quantum Hall effect—which describes a gapped topological phase—we are dealing with gapless states here. In such cases, a high wavefunction overlap does not necessarily imply that the states belong to the same universality class. Therefore, it is crucial to further verify whether the ansatz reproduces the correct scaling dimensions of the corresponding CFT.

\begin{table}[htbp]
    \caption{\label{tab:overlap} Wavefunction overlap $\langle \psi_{\textrm{ansatz}} | \psi_{\textrm{CFT}} \rangle$ between the proposed ansatz and the CFT ground state on the fuzzy sphere at different systems sizes $N=2s+1$. The ansatz takes the form of Eq.~\eqref{eq:ansatz_free} for the free scalar CFT and Eq.~\eqref{eq:ansatz_Ising} for the Ising CFT. The Ising CFT ground state is obtained using the same Hamiltonian as in Ref.~\cite{ZHHHH2022}.}
\setlength{\tabcolsep}{0.2cm}
\renewcommand{\arraystretch}{1.4}
    \centering
    \begin{tabular}{cccccccc}
        \hline\hline
& $N=12$ & $N=16$ & $N=20$ & $N=24$ & $N=28$ & $N=32$\\ 
Free CFT & 0.9930 & 0.9905 & 0.9879 & 0.9854 & 0.9829 & 0.9804  \\ 
Ising CFT & 0.9967 & 0.9951 & 0.9935 & 0.9918 & 0.9902 & 0.9885  \\
        \hline \hline
    \end{tabular}

\end{table}

One way to extract scaling dimensions from the ground state is through correlators. Specifically, we compute the two-point functions $\langle n^x(\bm x_1)\, n^x(\bm x_2) \rangle$ and $\langle n^z(\bm x_1)\, n^z(\bm x_2) \rangle - \langle n^z(\bm x) \rangle^2$, which will give the scaling dimensions of $\phi$, $\phi^2$ for the free scalar CFT, and $\sigma$, $\epsilon$ for the Ising CFT, respectively. In the infinite-$s$ limit, the expected CFT correlator behaves as
\begin{equation}
\frac{1}{\left( \sqrt{2s+1}\, \sin(\gamma_{12}/2) \right)^{2\Delta}} = \frac{1}{(2s+1)^\Delta} \sum_{n=0}^\infty \frac{(\Delta)_n}{n!} \cos^{2n}\left( \frac{\gamma_{12}}{2} \right),
\end{equation}
where $\gamma_{12}$ is the angle between $\bm x_1$ and $\bm x_2$, and $(\Delta)_n$ denotes the Pochhammer symbol. On the fuzzy sphere, the correlator takes the form of a truncated series in $\cos^2(\gamma_{12}/2)$, terminating at order $2s$:
\begin{equation}
\langle n^A(\bm x_1)\, n^A(\bm x_2) \rangle = \sum_{n=0}^{2s} a_n \cos^{2n}\left( \frac{\gamma_{12}}{2} \right).
\end{equation}
This allows us to extract the scaling dimension $\Delta$ either by fitting the antipodal correlator (i.e., at $\gamma_{12} = \pi$) as a function of $2s+1$, or by comparing the ratio $a_n/a_0$ with the CFT prediction $(\Delta)_n / n!$. 

\begin{figure}
    \centering
    \includegraphics[width=0.42\linewidth]{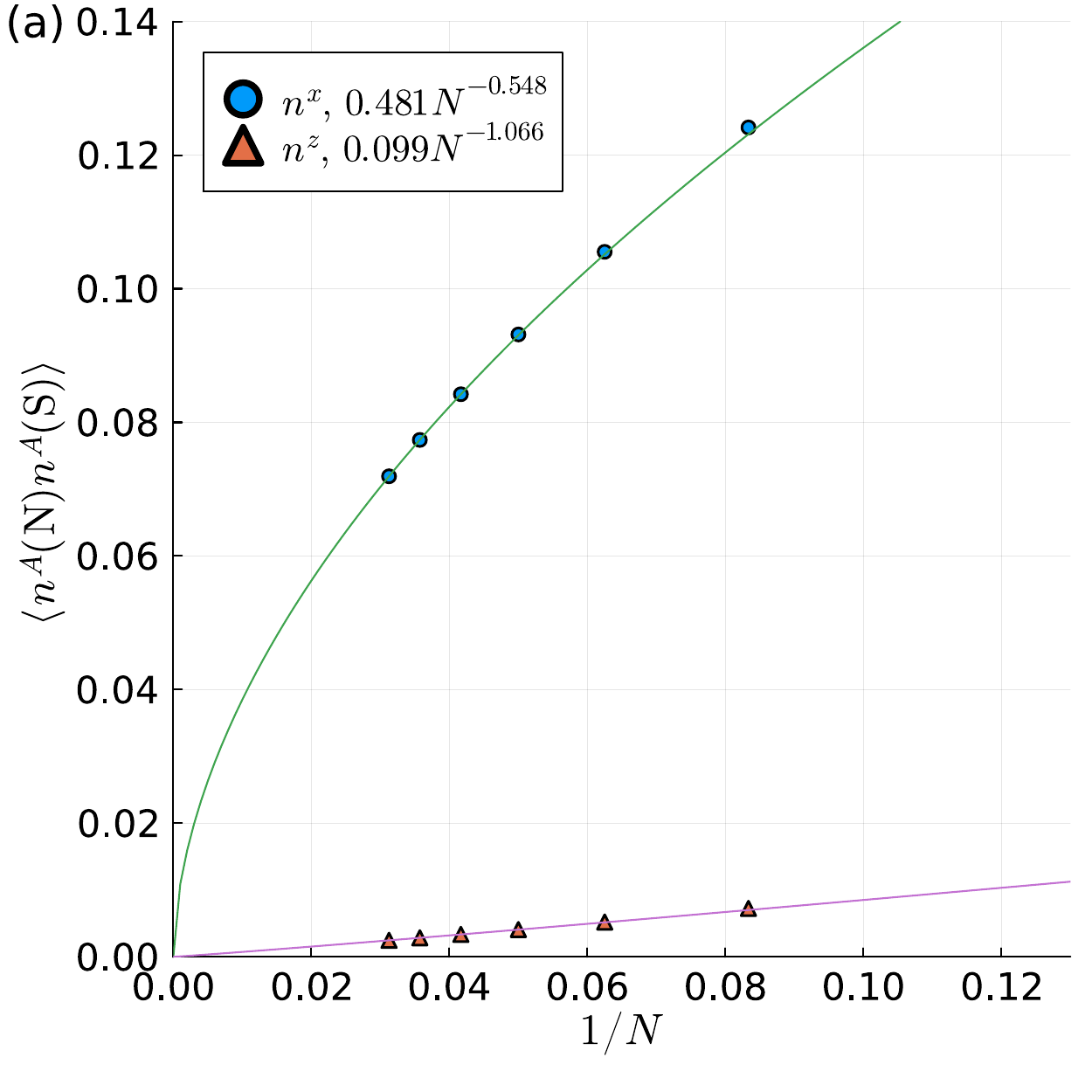}
    \includegraphics[width=0.28\linewidth]{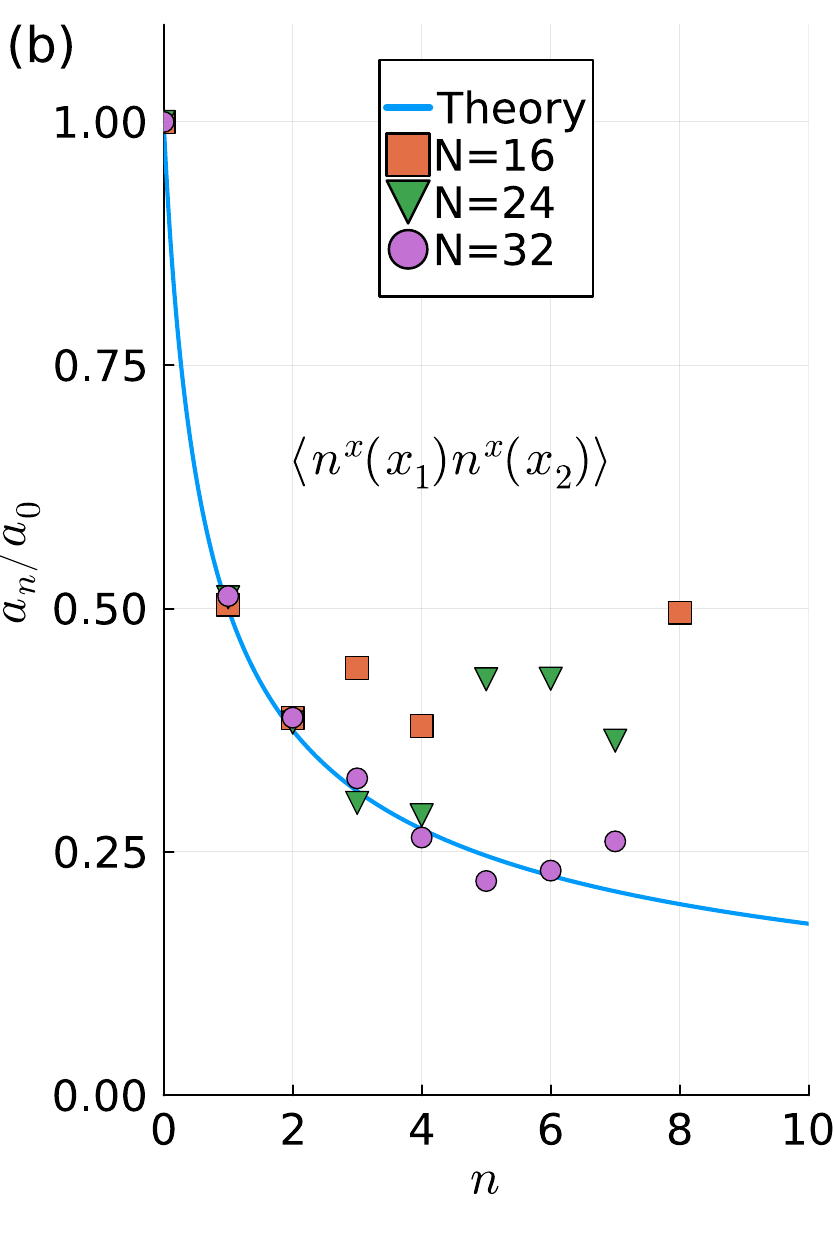}
    \includegraphics[width=0.28\linewidth]{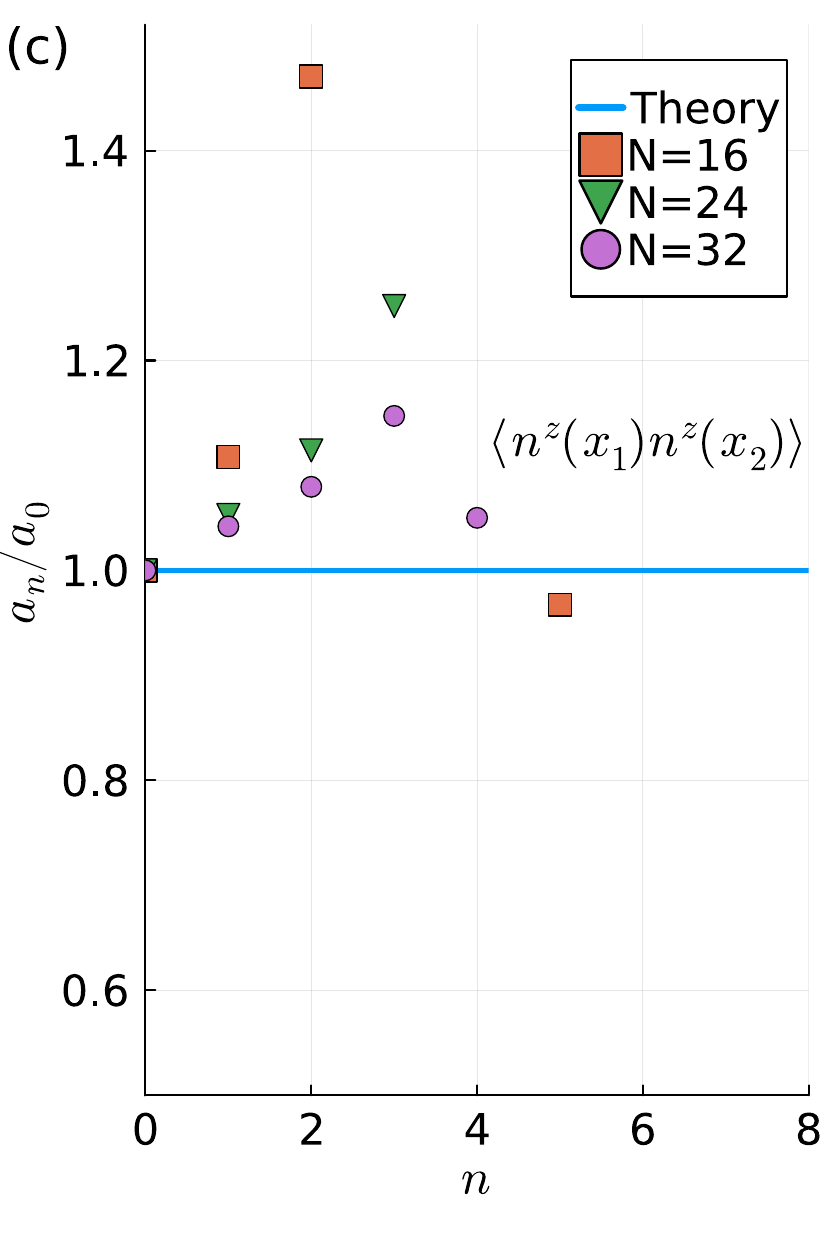}    
    \caption{Correlators of the free scalar  CFT ansatz: (a) Anti-podal correlator. (b), (c) Coefficients of series expansion of correlators $\langle n^x(\bm x_1) n^x (\bm x_2)\rangle$ and $\langle n^z(\bm x_1) n^z (\bm x_2)\rangle - \langle n^z(\bm x_1)\rangle^2$.}
    \label{fig:freeanstz_correlator}
\end{figure}

\begin{figure}
    \centering
    \includegraphics[width=0.42\linewidth]{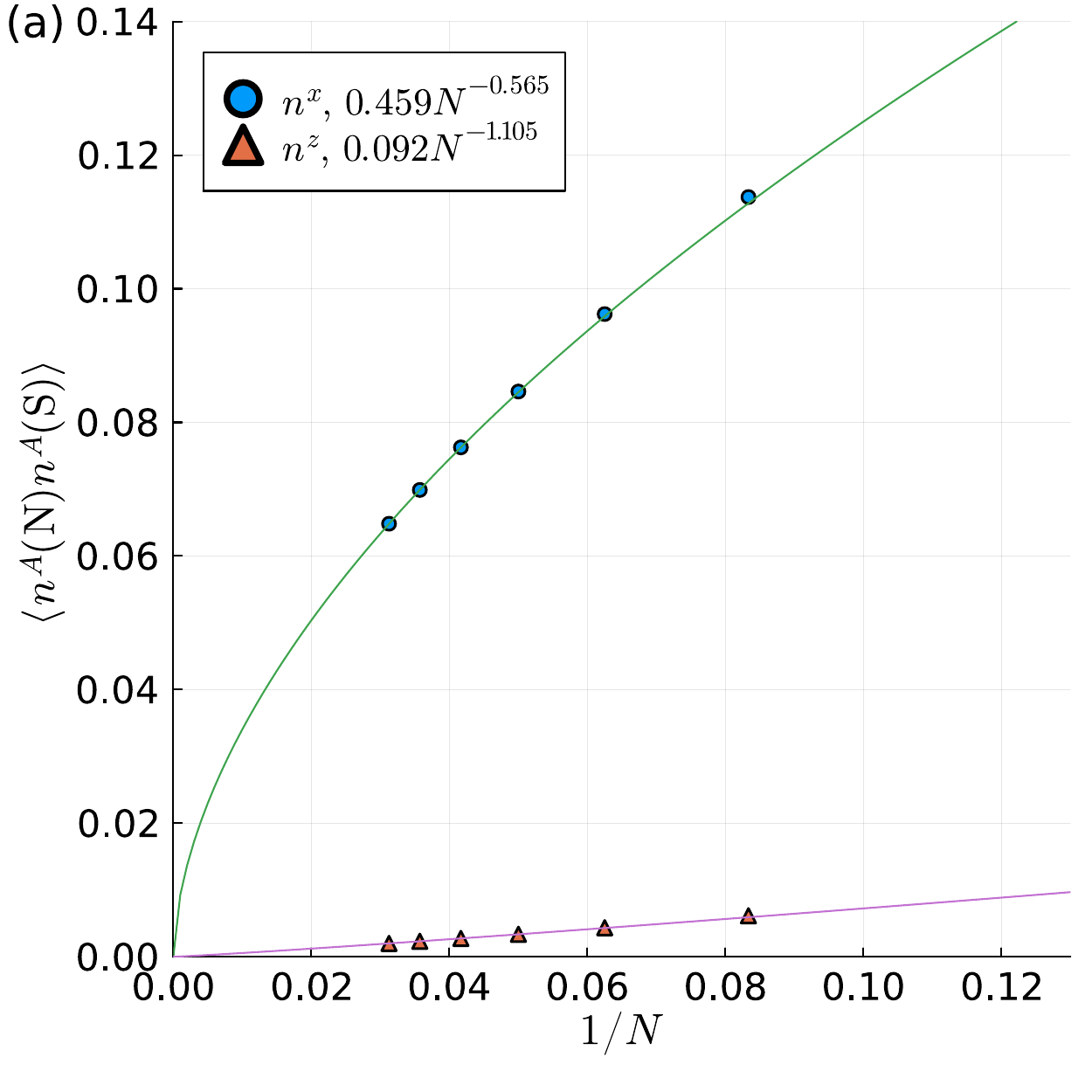}
    \includegraphics[width=0.28\linewidth]{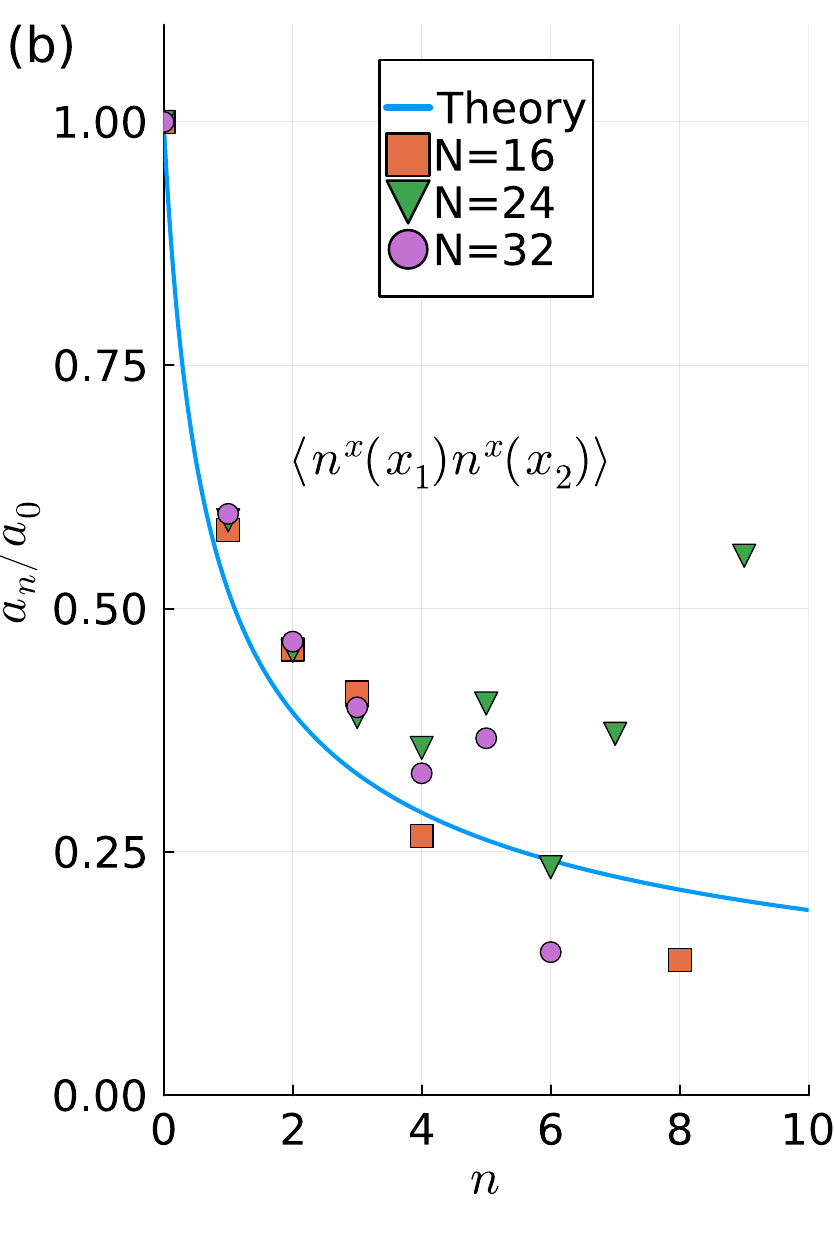}
    \includegraphics[width=0.28\linewidth]{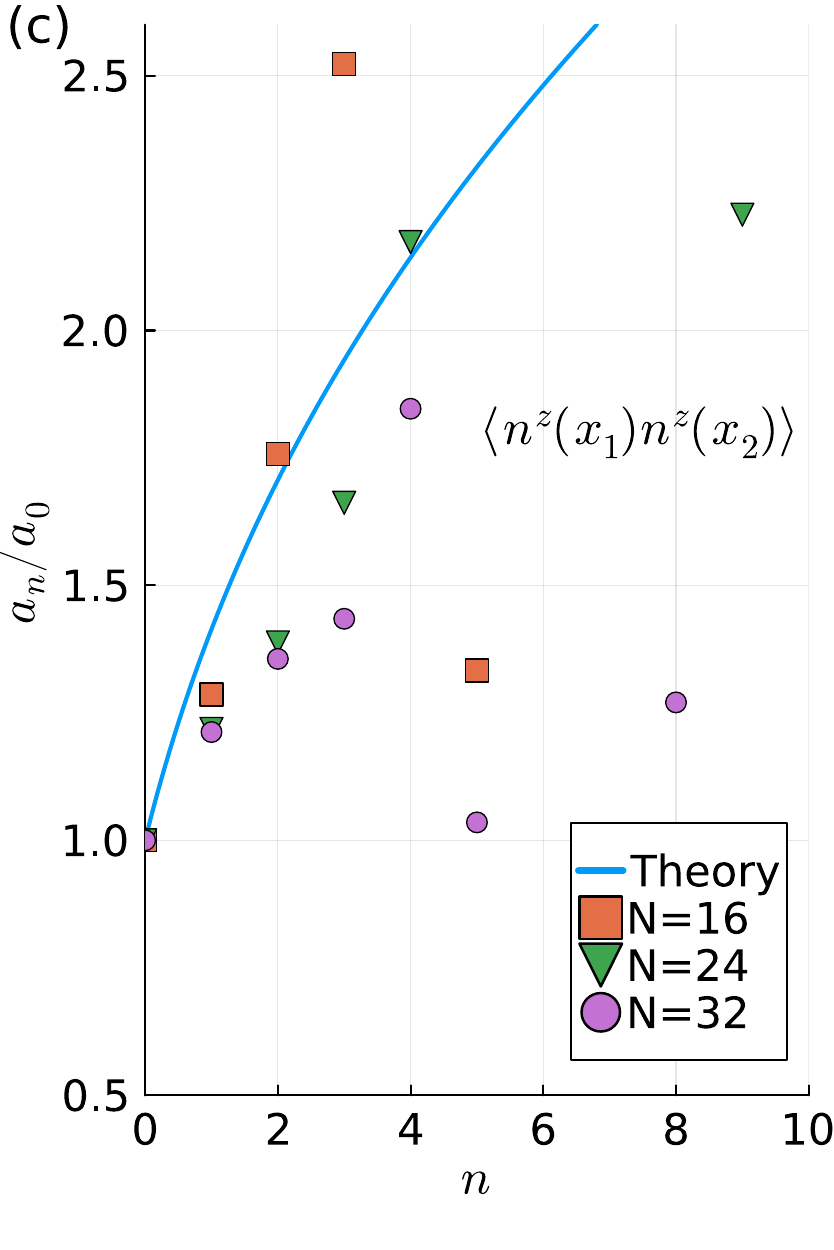}    
    \caption{Correlators of the Ising  CFT ansatz: (a) Anti-podal correlator. (b), (c) Coefficients of series expansion of correlators $\langle n^x(\bm x_1) n^x (\bm x_2)\rangle$ and $\langle n^z(\bm x_1) n^z (\bm x_2)\rangle - \langle n^z(\bm x_1)\rangle^2$.}
    \label{fig:Isinganstz_correlator}
\end{figure}

Fig.~\ref{fig:freeanstz_correlator}(a) shows the antipodal correlators of $n^x$ and $n^z$, whose fits yield scaling dimensions consistent with those of the free scalar CFT. Furthermore, as shown in Fig.~\ref{fig:freeanstz_correlator}(b) and (c), the ratios $a_n/a_0$ for the $n^x$ and $n^z$ correlators at small $n$ also align well with the theoretical expectations. The $n^x$ correlator exhibits excellent agreement with the predicted behavior, while the $n^z$ correlator shows somewhat larger discrepancies. Nevertheless, these discrepancies diminish as the system size increases, suggesting convergence to the correct scaling behavior in the continuum limit. In contrast, the correlators computed from the Ising CFT ansatz, shown in Fig.~\ref{fig:Isinganstz_correlator}, exhibit significantly larger deviations from the theoretical scaling dimensions $\Delta_{\sigma} = 0.518148806(24)$ and $\Delta_{\epsilon} = 1.41262528(29)$~\cite{bootstrapping3disingstress2025}. Notably, the discrepancies appear to grow with increasing system size. This trend may indicate that the proposed ansatz does not accurately capture the correct universality class of the 3D Ising CFT.

Our wavefunction ansatz performs a small unitary rotation on the trivial product state $|\varphi_0\rangle$, so its dominant component remains the unentangled product state. This structure explains several peculiar observations of the 3D Ising CFT on the fuzzy sphere. For instance, the ground state wavefunction exhibits very low double occupancy on each orbital, and the corresponding entanglement entropy is also notably small. Excited states may be constructed in a similar manner. For example, the Ising $|\sigma\rangle$ state can be approximated by applying a unitary rotation to $n^+_{\ell=0,m=0} |\varphi_0\rangle$, which carries a larger entanglement entropy. This perspective naturally explains the entanglement hierarchy observed among the low-lying states of the Ising CFT on the fuzzy sphere. Understanding the excited state wavefunctions may also shed light on potential algebraic structures underlying 3D CFT states—serving as a possible analogue of the Virasoro algebra in 2D CFTs.

It would be interesting to further investigate whether the 3D Ising CFT can be captured by the wavefunction ansatz in Eq.~\eqref{eq:ansatz_Ising} under suitable modifications—for example, by adjusting the numerical prefactors, such as replacing $\log\left( \frac{2(1-\Delta)_{\ell+1}}{(\Delta)_\ell} K_\ell \right)$ with an alternative functional form. It is already encouraging that our wavefunction ansatz achieves a high overlap with the Ising CFT ground state and, more importantly, exhibits power-law decaying correlators. However, it remains unclear whether this wavefunction corresponds to the ground state of a local Hamiltonian. Understanding how locality is encoded in the wavefunction is crucial. Imposing locality constraints may further restrict the form of the ansatz and could ultimately guide us toward a correct wavefunction for the 3D Ising CFT.

\section{Acknowledgment}

We thank Chong Wang, Zechuan Zheng, Zheng Zhou and Wei Zhu for illuminating discussions. We are especially grateful to Davide Gaiotto for numerous insightful discussions during the early stages of this project. Research at Perimeter Institute is supported in part by the Government of Canada through the Department of Innovation, Science and Industry Canada and by the Province of Ontario through the Ministry of Colleges and Universities. Part of the numerical calculations are done using the package FuzzifiED devoleped by Zheng Zhou~\cite{zhou2025fuzzifiedjuliapackage}.

\appendix 

\section{Cost function} \label{sec:cost} 

In this section, we explain the definition of the cost function. We define the cost function to capture the deviations of our scaling dimensions from the state-operator correspondence. The states we choose are $|\phi\rangle$, $|\partial_\mu\phi\rangle$, $|\partial_{\mu}\partial_{\nu}\phi\rangle$, $|\phi^2\rangle$,  $|\partial_\mu\phi^2\rangle$, $|\partial_{\mu} \partial_{\nu}\phi^2\rangle$ and $T_{\mu\nu}$. We define the vectors of energy and scaling dimension differences 
\begin{align}
    \mathbf{E}&=(E_{\phi}, E_{\partial \phi},E_{\partial\partial\phi},E_{\phi^2}, E_{\partial \phi^2},E_{\partial\partial\phi^2}, E_{T_{\mu\nu}})\nonumber\\
    \bDelta&=(0.5, 1.5, 2.5, 1, 2, 3, 3).
\end{align}
For perfect real scalar CFT, these two vectors should be proportional, 
\begin{equation*}
    \bDelta=\alpha\mathbf{E}.
\end{equation*}
The cost function is defined by their relative difference
\begin{equation}
Q(\alpha)= \sqrt{\frac{1}{ 7}\sum_{i=1}^7 ( 1 - \alpha \frac{E_i}{\Delta_i})^2}
\end{equation}
Here $\alpha$ is chosen to minimize $Q(\alpha)$ and can be evaluated directly
\begin{equation}
\alpha = \frac{\sum_i E_i/\Delta_i}{\sum_i E_i^2/\Delta_i^2}.
\end{equation}

\section{Determining effective harmonic oscillators on fuzzy sphere}
\label{sec:semidefinite}
Technically, the pair $(f_\ell, g_\ell)$ in Eq.~\eqref{eq:effa} is obtained as the eigenvector corresponding to the smallest eigenvalue of a generalized eigenvalue problem,
\begin{equation}
A \begin{pmatrix} f_\ell \\ g_\ell \end{pmatrix} = \lambda B \begin{pmatrix} f_\ell \\ g_\ell \end{pmatrix}, \nonumber
\end{equation}
where the matrices $A$ and $B$ are given by
\begin{equation}
A = \begin{pmatrix}
\langle n^x_{\ell,0} n^x_{\ell,0} \rangle & -i \langle n^y_{\ell,0} n^x_{\ell,0} \rangle \\
i \langle n^x_{\ell,0} n^y_{\ell,0} \rangle & \langle n^y_{\ell,0} n^y_{\ell,0} \rangle
\end{pmatrix}, \quad
B = \begin{pmatrix}
\langle n^x_{\ell,0} n^x_{\ell,0} \rangle & i \langle n^y_{\ell,0} n^x_{\ell,0} \rangle \\
- i \langle n^x_{\ell,0} n^y_{\ell,0} \rangle & \langle n^y_{\ell,0} n^y_{\ell,0} \rangle
\end{pmatrix}.
\end{equation}

For a generalized eigenvalue problem with
\begin{equation}
A = \begin{pmatrix} a & -b \\ -b & c \end{pmatrix}, \quad 
B = \begin{pmatrix} a & b \\ b & c \end{pmatrix},
\end{equation}
the eigenvector corresponding to the smallest eigenvalue is
\begin{equation}
(f, g) = \left( \sqrt{\frac{c}{2ac + 2b\sqrt{ac}}},\ \sqrt{\frac{a}{2ac + 2b\sqrt{ac}}} \right),
\end{equation}
such that
\begin{equation}
(f, g)\, A\, (f, g)^T = \frac{\sqrt{ac} - b}{\sqrt{ac} + b}, \quad 
(f, g)\, B\, (f, g)^T = 1.
\end{equation}
Note that both $A$ and $B$ are positive semidefinite, so the condition $ac \ge b^2$ holds. In the special case where $ac = b^2$, the solution reduces to
\begin{equation}
(f, g) = \frac{1}{2} \left( \frac{1}{\sqrt{a}},\ \frac{1}{\sqrt{c}} \right).
\end{equation}

\bibliography{ref}

\end{document}